\documentclass[twocolumn,prb,aps,longbibliography,superscriptaddress,nobalancelastpage]{revtex4-2}
\usepackage{times}
\usepackage{lineno}
\usepackage{amsmath,amsfonts,amssymb,graphicx,hyperref,epstopdf,xcolor,tikz,scalerel,natbib,array,gensymb}
\usepackage{textcomp,braket, physics, bbm, dsfont}
\usepackage[T1]{fontenc}
\usepackage[utf8]{inputenc}
\usepackage{multirow}
\usepackage{array}
\usepackage{soul}
\DeclareMathAlphabet{\mathpcal}{OMS}{zplm}{m}{n}

\graphicspath{{../figures/}}

\definecolor{lime}{HTML}{A6CE39}
\DeclareRobustCommand{\orcidicon}
{
	\begin{tikzpicture} 
	\draw[lime, fill=lime] (0,0) circle [radius=0.15] node[white] {{\fontfamily{qag}\selectfont \tiny ID}};
	\draw[white, fill=white] (-0.0625,0.095) 	circle [radius=0.007];
	\end{tikzpicture}
	\hspace{-2.5mm}
}
\newcommand\orcidID[1]{\href{https://orcid.org/#1}{\orcidicon}}

\newcommand{\be}{\begin {equation}}
\newcommand{\ee}{\end {equation}}
\newcommand{\beqa}{\begin {eqnarray}}
\newcommand{\eeqa}{\end {eqnarray}}

\hypersetup{colorlinks,citecolor=blue,filecolor=black,linkcolor=blue,urlcolor=blue}

\begin{document}

%\title{Entanglement signatures of topological phases in a 2D gapless Bogoliubov-de Gennes superconductor}

%\title{Characterizing topological phases in a 2D gapless Bogoliubov-de Gennes superconductor from entanglement perspective}
%\title{Characterizing gapless topological phases in a DIII TSC using entanglement1: Investigation of gapless topological phases in a 2D TRITOPS using entanglement\\}
\title{Entanglement signatures of gapless topological phases in a $p$-wave superconductor}

%\title{Characterizing topological phases in a 2D gapless Bogoliubov-de Gennes \\superconductor from entanglement perspective}
%\title{Characterizing a 2D gapless spin-filtered superconductor from entanglement perspective}
\author{S. Srinidhi \orcidID{0009-0004-3387-4908}}
\email[E-mail: ]{s.srinidhi312@gmail.com}
\affiliation{Department of Physics, Birla Institute of Technology and Science Pilani, Pilani Campus, Vidya Vihar, Pilani, Rajasthan 
333031, India.}
\author{Shashi C. L. Srivastava\orcidID{0000-0003-2049-5226}}
\email[E-mail: ]{shashi@vecc.gov.in}
\affiliation{Variable Energy Cyclotron Centre, 1/AF Bidhannagar, Kolkata 700064, India}
\affiliation{Homi Bhabha National Institute, Training School Complex, Anushaktinagar, Mumbai 400094, India}
\author{Jayendra N. Bandyopadhyay\orcidID{0000-0002-0825-9370}}
\email[E-mail: ]{jnbandyo@gmail.com}
\affiliation{Department of Physics, Birla Institute of Technology and Science Pilani, Pilani Campus, Vidya Vihar, Pilani, Rajasthan 
333031, India.}

\begin{abstract}
We explore the gapless topological phases of a $p$-wave superconductor, probing its rich topologically ordered phases and underlying quantum phenomena. The topological order of the system is characterized by studying its entanglement properties. This study confirms the bulk-boundary correspondence in the entanglement spectrum, even without a full bulk gap. For contractible bipartitions, the entanglement entropy varies non-monotonically with the chemical potential, displaying pronounced peaks at points where the bulk gap closes and reopens, signaling topological quantum phase transitions. This behavior remains robust in the thermodynamic limit. The entanglement entropy grows with system size for non-contractible bipartitions, indicating long-range entanglement in the gapless phase. These findings reveal the subtle interplay between symmetry, entanglement, and topology in gapless systems, and emphasize the role of entanglement-based diagnostics in identifying unconventional topological phases beyond the gapped paradigm.
\end{abstract}

\maketitle
\section{Introduction}
\label{Introduction}

Topological phases of matter, defined by global invariants and robust degenerate edges, have been extensively studied in gapped systems such as topological insulators and superconductors    \cite{TI_and_TSC_2011, Zoo_TQPT}. These phases are typically understood through bulk-boundary correspondence, where a finite energy gap in the bulk ensures the presence of non-trivial edge modes. However, gapless systems present a more subtle band spectrum. In the absence of a full bulk gap, the classification of topological phases becomes more complex, and the signatures of topological order are often masked by low-energy excitations    \cite{GSPT_2017, Intrinsic_GPT, Floquet_GPT, II_order_TI_SC}. This is due to nodal quasiparticles, which emerge in these gapless systems and dominate the low-energy excitations \cite{Nodal_Sc, Nodal_Qubit}. Despite this, certain gapless systems can still host nontrivial topological features, motivating the search for alternative probes beyond conventional topological invariants    \cite{PhysRevLett.133.026601, 1DGPT_NH, TI_GPT,2DKitaev, zhang2019, Srinidhi_25}.

Topological quantum phase transitions (TQPTs), which occur at zero temperature when a system’s ground state changes qualitatively as a function of a tuning parameter, are fundamental to understanding changes in the topological phases    \cite{PhysRevB.106.134430, PhysRevA.106.022208, PhysRevB.107.094415}. Beyond conventional phase transitions marked by symmetry breaking, certain topological phase transitions involve changes in global properties. Local order parameters like bandgap, occupation number, etc., cannot capture these transitions. Detecting such transitions, especially without a full bulk gap, requires tools sensitive to subtle changes in quantum correlations    \cite{Beyond_TPT_1, Corr_QPT, TQPT_RDM, RevModPhys.93.045003, TQPT_ML}.

Quantum entanglement has proven to be a powerful tool in this area of study    \cite{Peschel_2009, Eisler_2014, PhysRevB.89.104303, Eisler_2020,Eisler_2022}. In particular, the entanglement spectrum (ES), which is the set of negative $\log$ of eigenvalues of the reduced density matrix (RDM), encodes information analogous to edge-state spectra and serves as a sensitive diagnostic tool of topological order  \cite{Fidkowski, Aditi, QSL_chiral, PhysRevB.87.045115,Eisler_2009}. The characteristic spectral structures, like the ground state degeneracy and long-range entanglement, often reflect the topological order of the gapped system. While its role is well established in gapped systems, the behavior of the ES in gapless systems is more intricate and less understood \cite{XXZ_EE, Gapless_spin_chain, 2D_gapless, COE_Shunshuke}. Investigating its structure in such regimes offers a valuable perspective on topological systems that lack a finite spectral band gap. In particular, ES derived from various types of bipartitions provides direct information of the underlying topology: (i) contractible bipartitions (CBs), where the system is partitioned into trivially connected regions, and (ii) a non-contractible bipartition (NCB), in which the cut wraps around the system's topology sector, offering complementary insight about the topological order of the system \cite{JHEP_2018, HAMMA200522}. While CB subsystems often highlight local entanglement features and can reveal bulk-boundary correspondences in gapped phases, NCB cuts are more sensitive to global topological order and long-range entanglement. Collectively, this study provides a clearer signature of TQPTs, even in complex or gapless systems \cite{Hamma_lattice_spin, Grover_2013}. Bipartite entanglement captures the essential physics, making it a natural choice for investigation, without exploring multipartite entanglement to characterize the non-interacting topological systems. \cite{Peschel_2009, Eisler_2009, Grover_2013}.

Furthermore, entanglement entropy (EE) reflects the topology of the system through its scaling behavior and subleading corrections, which are directly linked to the underlying pattern of quantum correlations \cite{2D_geometric_entanglement, fphy, Scaling_EE, EE_probe_AAH}. For gapped phases, it typically follows an area law with subleading topological corrections   \cite{Scaling_EE, EE_fractalization, Pirmoradian2024, SciPostPhys.17.1.010}. In contrast, gapless systems may exhibit nontrivial scaling and sensitivity to microscopic parameters, such as the chemical potential    \cite{2D_gapless, COE_Shunshuke}. Their derivative, with respect to a system parameter, often serves as a sensitive probe for TQPTs. In particular, peaks or discontinuities in the derivative signal quantum critical points (QCPs) that cause TQPTs, which are otherwise difficult to detect through spectral measures alone    \cite{Kitaev_Ladder, Critical_Free_fermion, Hamma_lattice_spin}.

In this work, we investigate a two-dimensional (2D) gapless $p$-wave superconductor (SC) described by the BdG formalism that preserves time-reversal symmetry (TRS). Such a system belongs to the topological class DIII with unconventional ``$p_x + p_y$'' type pairing potential \cite{TRI_review,TRI_1,TRI_2,TRI_3}. The $p$-wave superconductor in this class is characterized by helical Majorana edge modes protected by TRS and classified by a $\mathbb{Z}_2$ topological invariant. Unlike fully gapped topological superconductors, $p$-wave SC can host gapless excitations in the bulk while maintaining robust topological features at the boundaries \cite {2DKitaev, zhang2019, Srinidhi_25}. We analyze the system's topological phases and transitions using the entanglement spectrum and the sensitive dependence of entropy on the chemical potential. We consider both contractible and non-contractible bipartitions, revealing distinct entanglement features that reflect the presence of long-range correlations. Our results demonstrate that entanglement-based diagnostics remain robust despite the absence of a spectral gap, and provide a broader framework for identifying and 
characterizing unconventional topological phases.

Additionally, our prototype model provides valuable insights into the entanglement structure of time-reversal invariant topological superconductors (TRITOPS) \cite{TRI_1,TRI_2,TRI_3}, which are of increasing significance in the field of quantum computing. Combining two copies of our model with appropriate parameters is equivalent to a description of TRITOPS, expanding its relevance to topological quantum computing \cite{TRI_review}, where such systems are often proposed as a platform for fault-tolerant quantum information processing. These systems, with their ability to support robust edge states, are of particular interest for the development of topologically protected quantum bits (or qubits) \cite{QC_Majorana_Kramers}. Studying the gapless phases in such systems is crucial for understanding the dynamics of these qubits, as gapless topological superconductors exhibit unique responses that can influence quantum computation protocols \cite{GSPT_Bernevig, Kondo_anyons_TQC}. Moreover, the realizability of our prototype model makes it highly relevant for experimental efforts. Hybrid structures, for example, combining spin-spiral magnetic textures with conventional s-wave superconductors, have already demonstrated signatures consistent with emergent ($p_x + p_y$)-type pairing and gapless topological phases \cite{exp_realisation_Arijit}. This connection to real-world systems underscores the importance of studying such models in detail in the ongoing pursuit of novel quantum materials and technologies.

The remainder of this paper is structured as follows: In Sec. \ref{Theory and Techniques}, we introduce the model and outline the numerical methods. Section \ref{Results_CB} focuses on the entanglement properties under contractible bipartitions, while Sec. \ref{Results_NCB} addresses non-contractible cuts. In Sec. \ref{ES to QPT}, we analyze the behavior of entanglement near the topological quantum phase transitions (TQPTs), using the derivative of the entanglement entropy with chemical potential, $\left(\frac{dS_\Omega}{d\mu }\right)$ and the entanglement spectrum (ES). We conclude in Sec. \ref{Conclusion} with a summary of our findings and a discussion of potential experimental realizations and future directions. The Appendix \ref{app}, \ref{app_symm} is organized in parallel with the main text and elaborates on characterizing the topological phases undergoing phase transition and a proof of the existence of topological defects in the model arising from nodal superconducting band touching points. We provide details about the symmetry class of p-wave superconductors in appendix \ref{app_symm}.

\section{Theory and Techniques}
\label{Theory and Techniques}

\subsection{Model}
\label{Model}

\begin{figure}[t]
\includegraphics[width=8.5cm, height=6cm]{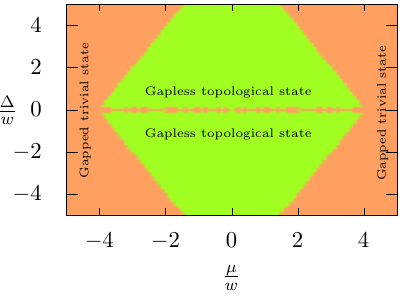}
\caption{The phase diagram of the $p$-wave SC on the $\mu - \Delta$ parameter plane (in units of $w$). The green region indicates the gapless topological state of the model with protected edge states, which separates it from the orange region (gapped trivial states). The corresponding band structure varying $\mu$ is presented in Fig. \ref{Energy_spectrum}. The system parameters are varied to depict the topology of the model in Eq. \eqref{h_real}.}
\label{Phase}
\end{figure}
We study the spinless $p$-wave SC on a square lattice, where nearest-neighbor (NN) hopping and pairing interactions occur along the $x$- and $y$-directions. To preserve TRS, we consider a ``$p_x + p_y$'' type superconducting pairing, which is kept equal along both spatial directions. This pairing symmetry preserves TRS and allows for nontrivial nodal structures in the Brillouin zone, which is crucial for realizing topological gapless phases. The real-space Hamiltonian for the model is given by:
\begin{align}
H &= \mu \sum_{n_x, n_y} c^\dagger_{n_x, n_y} c_{n_x, n_y} \notag \\ 
&\quad + w \sum_{n_x, n_y} \left( c^\dagger_{n_x, n_y} c_{n_x+1, n_y} 
+ c^\dagger_{n_x, n_y} c_{n_x, n_y+1} + \text{H.c.} \right) \notag \\ 
&\quad + \Delta \sum_{n_x, n_y} \left( c^\dagger_{n_x, n_y} c^\dagger_{n_x+1, n_y} 
+ c^\dagger_{n_x, n_y} c^\dagger_{n_x, n_y+1} + \text{H.c.} \right).
\label{h_real}
\end{align}
Here, the operators $c_{n_x, n_y}^\dag (c_{n_x, n_y})$ create (annihilate) a fermion at site $(n_x, n_y)$. The parameters $\mu$, $w$, and $\Delta$ represent the chemical potential, nearest-neighbor hopping potential, and the pairing potential, respectively. We assume equal hopping and pairing strengths along both lattice directions.  Being a quadratic model, the Hamiltonian can be written as $H = \Psi^\dagger \mathcal{H}_{\text{BdG}} \Psi,$ with the Nambu spinor $\Psi$, defined as,
\begin{align}
\Psi = \begin{aligned}
&\Big[ c_{1,1}, c_{1,2}, \dots, c_{1,N_y}, \;
c_{2,1}, \dots,c_{2,N_y}, \dots, c_{N_x,N_y}, \\ 
&\quad c^\dagger_{1,1}, c^\dagger_{1,2}, \dots, c^\dagger_{1,N_y}, \;
c^\dagger_{2,1}, \dots, c_{2,N_y}^\dagger, \dots,c^\dagger_{N_x,N_y} 
\Big]^{\mathrm{T}}.
\end{aligned}
\label{e_vec}
\end{align}
This doubled Nambu spinor representation, referred to as the BdG basis representation, explicitly includes particle and anti-particle degrees of freedom \cite{Kitaev, LSM}. Using periodic boundary conditions, the momentum space Hamiltonian is obtained by the Fourier transform,
\begin{subequations}
\be
c_{n_x, n_y} = \frac{1}{\sqrt{N}} \sum_{\mathbf{k}} e^{i \mathbf{k} \cdot \mathbf{r}} c_{\mathbf{k}}
\ee
\be
H = \sum_{\mathbf{k}} \Psi^\dagger_{\mathbf{k}} \, \mathcal{H}(\mathbf{k}) \, \Psi_{\mathbf{k}}, 
\ee
where the Nambu spinor and the $\mathcal{H}(\mathbf{k})$ are given as,
\be
\Psi_{\mathbf{k}} = \begin{pmatrix}
c_{\mathbf{-k}} &
c^\dagger_{\mathbf{k}}
\end{pmatrix} ^\mathrm{T}.
\ee
\be
\mathcal{H}(\mathbf{k}) = \epsilon(\mathbf{k}) \tau_z + \chi(\mathbf{k})\tau_y 
\label{hk}
\ee
\begin{align}
\text{where, }&\epsilon(\mathbf{k}) = \mu + 2w [\cos(k_x) + 
\cos(k_y)],\notag\\&  \chi(\mathbf{k}) = 2i\Delta [\sin(k_x) + \sin(k_y)].
\end{align}
\label{H_kspace}
\end{subequations}
\begin{figure}[t]
\includegraphics[width=8.5cm, height=5.8cm]{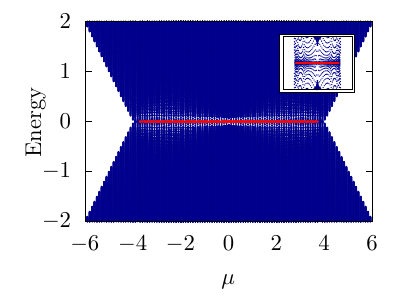}
\caption{Energy spectrum of the $p$-wave SC described in Eq. \eqref{h_real} with varying $\mu$. The inset plot depicts the band spectrum in the energy range $[-0.1:0.1]$. The other parameters are $w=1$, $\Delta=0.8w$, and $N_x=N_y=100$. The blue region denotes the gapless bulk modes, and the red data points denote the localized Majorana zero modes.}
\label{Energy_spectrum}
\end{figure}
Here, $\tau_i$ are the pseudospin matrices acting on the quasiparticle (Nambu) basis. The model preserves all three fundamental symmetries: time-reversal, chiral, and particle-hole symmetries, and belongs to the DIII topological class. The Appendix \ref{app_symm} discusses the topological classification in detail with additional information that bridges our prototype model to the TRITOPS \cite{TRI_review}.

The phase diagram shown in Fig. \ref{Phase} illustrates the topological behavior of the model in the $\mu$-$\Delta$ plane \cite{zhang2019, 2DKitaev}. The green region corresponds to a gapless topological phase, where the system supports robust edge states despite the absence of a full bulk gap. In contrast, the orange region corresponds to a trivial gapped phase. This distinction becomes evident in the energy spectrum shown in Fig. \ref{Energy_spectrum}, where we fix the $p$-wave pairing amplitude to $\Delta = 0.8w$, a choice that will be maintained throughout this work. The symmetry classification of the model and the topological phase characterization are discussed in detail in the Appendices.

From Fig. \ref{Energy_spectrum}, we observe that for chemical potentials in the range $-4w < \mu < 4w$, the spectrum features zero energy states localized at the system’s boundaries, indicative of topological edge modes. These edge states are protected by the underlying topological structure of the system despite its gapless nature \cite{zhang2019, Srinidhi_25}. This behavior is a hallmark of a topological quantum phase transition (TQPT), where the system evolves from a trivial gapped phase into a gapless topological one as $\mu$ is tuned across the critical values $\mu_c = \pm 4w$. The gapless nature of the spectrum arises from the simultaneous vanishing of both the kinetic and pairing terms, i.e., when $\epsilon(\mathbf{k}) = 0$ and $\chi(\mathbf{k}) = 0$. The pairing amplitude $\chi(\mathbf{k})$ vanishes along the lines $k_y = -k_x$ in the Brillouin zone, while $\epsilon(\mathbf{k})=0$ defines contours dependent on the chemical potential $\mu$. Consequently, gap closings occur at discrete points where these conditions coincide. TQPTs are marked by bulk gap closings at critical chemical potentials, $\mu_c = \pm 4w$, where the system transitions from a gapless topological phase into a fully gapped trivial phase. The system remains gapless for chemical potentials within this range, exhibiting multiple gap closings and openings associated with complex nodal structures and topological properties. Beyond $|\mu_c|$, the bulk gap reopens fully, and the system enters a trivial gapped phase. Thus, by tuning $\mu$, the model exhibits a rich phase diagram comprising an extended gapless topological phase bounded by trivial gapped phases beyond $\mu = \pm 4w$ [refer to Figs. \eqref{Phase} and \eqref{Energy_spectrum}]. Appendix \ref{app_defects} provides a detailed account of the nodal band touchings, including analytical derivations, and their implications for the low-energy spectrum.

\subsection{Entanglement-based TQPT characterization techniques}
\subsubsection{Entanglement Entropy}
\label{RDM}
\begin{figure}
\includegraphics[width=1\columnwidth]{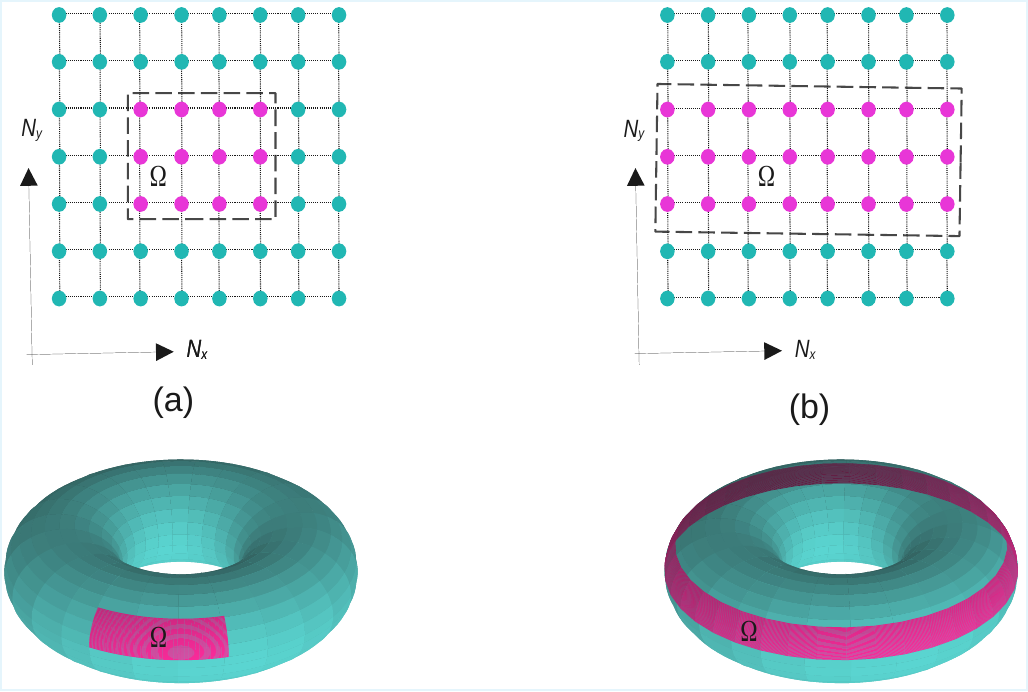}
\caption{The figure demonstrates the different types of bipartitions discussed 
in Sec. \ref{Results_CB} and \ref{Results_NCB}. Panel (a) demonstrates the 
contractible bipartition. The subsystem lattice parameters are set at 
$\left(N_x, N_y\right) = 40 \times 40$. Panel (b) demonstrates the 
non-contractible bipartition.  Here, the subsystem lattice sites are set to be  
$\left(N_x, N_y\right) = 40 \times N_y$ of the model Hamiltonian. The red region 
denotes the subsystem $\Omega$, and the rest of the pink region in the torus 
denotes the $\bar{\Omega}$. The figure is depicted for $N_x=N_y=N=80$. }
\label{CB_NCB_diag}
\end{figure}

To investigate the gapless topological phases through entanglement 
measures, we employ the fact that the reduced density matrix (RDM) can be obtained from the correlation function of the ground state of BdG 
Hamiltonian $\mathcal{H}_{\text{BdG}}$ defined in Eq. \eqref{h_real}. Following the Lieb-Schultz-Mattis (LSM) formalism    \cite{LSM}, the Hamiltonian can be rewritten as a quadratic form involving two matrices $A$ (Hermitian) and $B$ (anti-Hermitian):
\be
H = \sum_{i,j} c_{i}^\dagger A_{i,j} c_{j} 
+ \frac{1}{2} \sum_{i,j} \left( c_{i}^\dagger B_{i,j} c_{j}^\dagger - c_{i} B_{i,j}^* c_{j} \right),
\label{LSM}
\ee
where, $i = (i_x, i_y)$ and $j = (j_x, j_y)$ denote the two-dimensional lattice site indices. For computational convenience, these 2D indices are flattened into single indices $i, j = 1, \dots, M$ with $M = N_x \times N_y$, preserving a one-to-one correspondence between the 2D lattice and the matrix representation    \cite{NJP}. Applying the Bogoliubov transformation diagonalizes the Hamiltonian into quasiparticle modes:
\be
H = \sum_{\alpha = 1}^{M} \xi_\alpha \left( \eta_\alpha^\dag \eta_\alpha - \frac{1}{2} \right).
\ee
Here, the quasiparticle operators $\eta_\alpha$ are given by,
\be
\eta_\alpha = \sum_{i} \frac{1}{2} \Big( [\Phi_\alpha(i) + \Psi_\alpha(i)] c_i + [\Phi_\alpha(i) - \Psi_\alpha(i)] c_i^\dag \Big).
\ee
The normalized eigenvectors $\Phi_\alpha, \Psi_\alpha$ and corresponding energies $\xi_\alpha$ are solutions of the coupled equations:
\begin{align}
(A - B) \Phi_\alpha &= \xi_\alpha \Psi_\alpha, \\
(A + B) \Psi_\alpha &= \xi_\alpha \Phi_\alpha,
\end{align}
which can be combined into the eigenvalue problems    \cite{NJP}:
\begin{align}
(A - B)(A + B) \Psi_\alpha &= \xi_\alpha^2 \Psi_\alpha, \\
(A + B)(A - B) \Phi_\alpha &= \xi_\alpha^2 \Phi_\alpha.
\end{align}
The correlation matrix $G$, capturing two-point correlations within the ground state $|\text{GS}\rangle$, is defined as
\be
G_{i,j} \equiv \langle \text{GS}| (c_i^\dag - c_i)(c_j^\dag + c_j) |\text{GS} \rangle,
\ee
where, the indices $i,j$ correspond to flattened 2D sites. Expressed in terms 
of the eigenvectors, the matrix elements are defined as:
\be
G_{i,j} = -\sum_{\alpha} \Psi_\alpha(i) \Phi_\alpha(j).
\ee
The subsystem matrix for our choice of subsystem $\Omega$ is depicted in the Fig. 
\ref{CB_NCB_diag}, 
\be
T_\Omega \equiv G_\Omega G_\Omega^T= G_{i,j} G_{i,j}^T \quad \forall \, \, i,j \in \Omega
\ee
Entanglement entropy (EE) quantifies the quantum correlations between two subsystems of a quantum system. For quadratic Hamiltonians, the entanglement structure is entirely captured by two-point correlation functions, notably the subsystem correlation matrix $G_\Omega$ defined in the above equation \cite{Peschel_2009, TQPT_RDM}. The eigenvalues of the reduced density matrix are obtained from the eigenvalues $\xi_i^2 \in [0,1]$ of the subsystem matrix $T_\Omega$ \cite{Peschel_2009, PhysRevB.89.104303}. The von Neumann entanglement entropy is then computed as a sum over binary Shannon entropies    \cite{Kitaev_Ladder}:
\be   
S_\Omega = - \sum_i \left\{ \left(\frac{1+\xi_i}{2}\right) \ln \left(\frac{1+\xi_i}{2}\right)+ \left(\frac{1-\xi_i}{2}\right) \ln \left(\frac{1-\xi_i}{2}\right) \right\}.
\ee
The reduced density matrix eigenvalues are given by $\frac{1 \pm \xi_i}{2}$, where $\xi_i$ are the square roots of the eigenvalues of the $T_\Omega$. Here, $\xi_i = 0$, the modes correspond to maximally entangled states, representing zero-energy modes (e.g., Majorana zero modes) that contribute maximal quantum correlations. On the other hand, $\xi_i=\pm 1$ corresponds to fully occupied or empty bulk modes. These modes contribute minimally to the entanglement entropy and thus represent minimally entangled states. Therefore, for BdG systems, the entanglement entropy contribution from a mode is maximal when $\xi_i= 0$ and minimal when $\xi_i=\pm 1$.

\subsubsection{Entanglement Spectrum}
\label{ES}

To analyze the quantum correlations in the system, we focus on the entanglement spectrum (ES), which provides a more refined characterization of entanglement than EE alone \cite{Fidkowski, Aditi, PhysRevLett.133.026601, QSL_chiral, Gapless_spin_chain}. The Li-Haldane conjecture states that the low-lying spectrum of the entanglement Hamiltonian, obtained from the RDM, reflects the physical edge-state spectrum of a system in a topological phase. In other words, the ES encodes universal information about the system's boundary excitations, serving as a powerful probe of topological order. This connection bridges entanglement properties with boundary physics in quantum many-body systems \cite{Li_Haldane_2008}. Starting from the correlation matrix $G$ for the full system, we restrict it to a bipartitioned subsystem $\Omega$ to obtain the subsystem correlation matrix $ G_\Omega$ \cite{Eisler_2009,Peschel_2009}. This matrix contains complete information about the subsystem matrix $T_\Omega$, which is connected with the mixed state of the subsystem when traced over its complement $\bar{\Omega}$ denoted in the Fig. \ref{CB_NCB_diag}  \cite{NJP, Peschel_2009, SciPostPhys.17.1.010}.

The ES is then computed by diagonalizing $ T_\Omega $, whose eigenvalues correspond to the squared singular values of $ G_\Omega $. These eigenvalues encode the spectrum of entanglement energies, offering a detailed fingerprint of how quantum correlations are distributed between the subsystems $\Omega$ and $\bar{\Omega} $. \cite{Fidkowski, Aditi, SciPostPhys.17.1.010}. The choice of the geometry of $\Omega$ decides the ES response    \cite{Grover_2013}, which is discussed in detail in the upcoming sections.

\section{Results}
\label{Results}

In quantum systems, particularly gapless cases, analyzing the entanglement properties under different bipartitions reveals key information about quantum correlations and criticality in the topology of the system. The nature of this division significantly influences the type and structure of information observed between the two subsystems  \cite{Grover_2013}. Understanding them is crucial for studying QCPs, topological effects, and the low-energy spectrum, which play an important role in gapless systems.

\subsection{Contractible bipartitioned system}
\label{Results_CB}
\begin{figure}
\includegraphics[width=1\columnwidth]{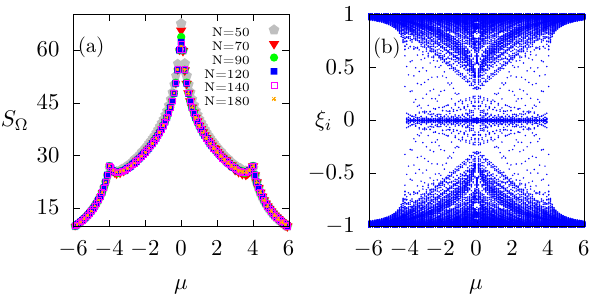}
\caption{Entanglement entropy (a) and the entanglement spectrum (b) for the contractible boundary cut with varying $\mu$. The lattice dimension is set to be $M=N \times N=80\times80$ for the ES in the panel (a) and varied for the EE in the panel (b). The dimension of $\Omega$ is given in Fig. \ref{CB_NCB_diag}.}
\label{CB_EE&ES}
\end{figure}

A contractible bipartition (CB) refers to a localized subsystem that can be continuously deformed and shrunk to a point without crossing the system boundary \cite{Grover_2013}. For instance, excising a patch-shaped region from the surface of a torus creates a contractible bipartition, as illustrated in Fig. \ref{CB_NCB_diag} (a). The subsystem dimension is chosen to be $M_\Omega = \left(  40 \times 40 \right) $, where $N_x=N_y=N$ are varied as per the dimension given in the Fig. \ref{CB_EE&ES}(a).

The entanglement entropy (EE) under CBs typically obeys an area law scaling with the boundary length between the subsystem and its environment in gapped systems. This reflects that quantum correlations are short-ranged and confined to the vicinity of the entanglement cut \cite{Scaling_EE, JHEP_2018, fphy,EE_probe_AAH}. Even in gapless systems, the EE under a CB remains sensitive primarily to local entanglement and does not exhibit volume scaling. Consequently, it serves as a useful local probe of quantum correlations. 

We begin our analysis using contractible bipartitions by studying the EE in the $p$-wave SC. The entropy is computed for fixed subsystem geometry across various system sizes $M = 50\times50, 70 \times70, 90\times90, 120\times120, 140 \times140, 180 \times 180$, and is depicted in Fig. \ref{CB_EE&ES} (a). The EE displays non-monotonic behavior as the chemical potential $\mu$ is varied. Notably, pronounced peaks appear at specific values of $\mu$ where the bulk energy gap closes and reopens, indicating TQPTs. These features are robust across different system sizes ($M$). Additionally, the entropy curves are symmetric about $\mu = 0$, consistent with the particle-hole symmetry inherent in the BdG form of the Kitaev Hamiltonian    \cite{Kitaev, zhang2019, 2DKitaev, Srinidhi_25}.

Although contractible subsystems do not capture non-local topological features or ground state degeneracies \cite{Grover_2013}, the EE still encodes valuable physical information. In particular, our results show that it acts as a diagnostic for criticality and gaplessness, revealing the presence of low-energy excitations even in the absence of a bulk gap. The prominent entropy peaks make it a sensitive probe of TQPTs from a local entanglement perspective.

Figure \ref{CB_EE&ES}(b) shows the ES for CBs as a function of chemical potential $\mu$. The spectrum displays a depletion of maximal entanglement modes as $\mu \to 0$.  In the region $\mu \in [-4,4]$, the ES shows a dense maximally entangled state at $\xi = 0$, signaling an edge-like behavior in the system. The ES is fully gapped for $\abs{\mu} > 4$, where the system is in a trivial state. This qualitative change signals a reorganization of entanglement structure, in agreement with the signature of TQPTs in the energy spectrum shown in Fig. \ref{Energy_spectrum}. Although these signatures do not directly infer to the topological order, they nevertheless hint at the underlying topological signatures encoded in the entanglement structure of the system.

\subsection{Non-contractible bipartitioned system}
\label{Results_NCB}

\begin{figure}
\includegraphics[width=1\columnwidth]{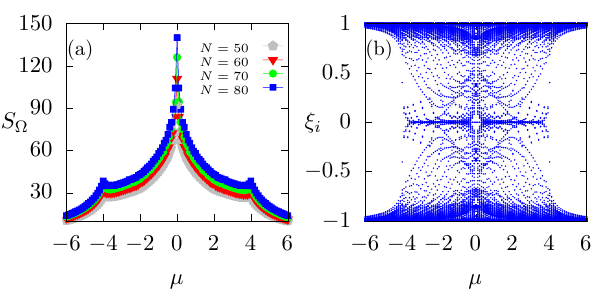}
\caption{Entanglement entropy (a) and the entanglement spectrum (b) for the non-contractible boundary cut with varying $\mu$. The lattice dimension $M= N \times N = 80 \times 80$ for the ES in the panel (a) and varied for the EE in the panel (b). The dimension of $\Omega$ is given in Fig. \ref{CB_NCB_diag}.}
\label{NCB_EE&ES}
\end{figure}
\begin{figure}[b]
\includegraphics[width=1\columnwidth]{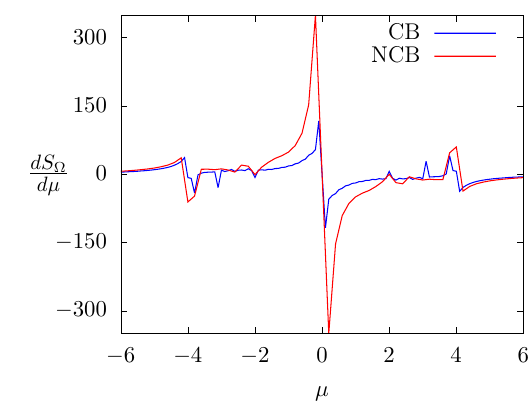}
\caption{The derivative of the entanglement entropy (EE) computed for both CB and NCB subsystems. The system parameters $\Delta = 0.8w$ and $N = 80$ are used for this analysis.}
\label{dEE}
\end{figure}
In contrast to CBs, a non-contractible bipartition (NCB) involves a cut that wraps around a non-trivial cycle of the underlying topology, such as the hole of a torus. This type of partition cannot be continuously deformed into a point [refer to Fig. \ref{CB_NCB_diag} (b)]. Instead, it probes the global properties of the system. The resulting subsystems are non-local, each intersecting a non-contractible loop of the manifold. Such cuts reveal long-range quantum correlations, often associated with topological order and criticality. In gapped topologically ordered systems, NCBs can directly indicate the ground state degeneracy \cite{Grover_2013}. Our results demonstrate that, even in gapless systems, they remain a powerful probe of non-local entanglement and topological features embedded in the low-energy structure.
\begin{figure*}
\includegraphics[width=1\textwidth]{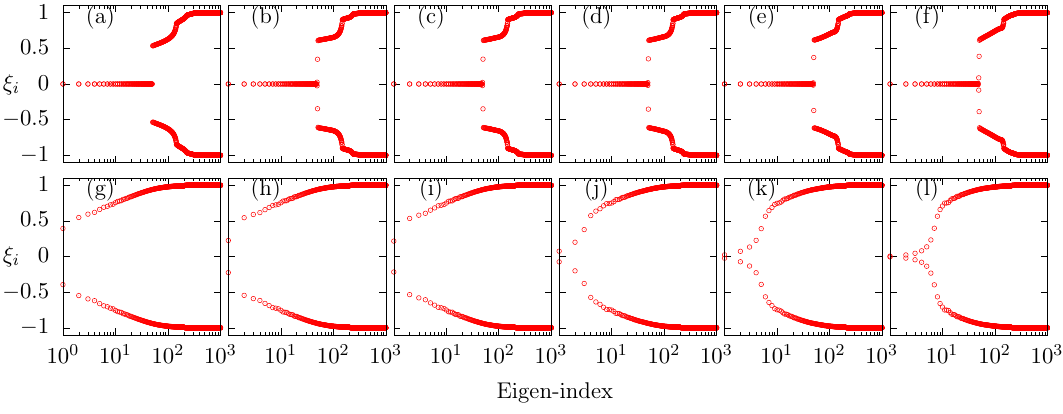}
\caption{The ES for a CB subsystem near the QCPs. Panels (a)-(f) refer to the TQPT happening around $\mu=0$, where  $\abs{\mu}>0$ is a topological phase. The chemical potential is chosen to be: (a) $0.0$, (b) $0.001$, (c) $0.005$, (d) $0.01$, (e)  $0.05$, and (f) $0.1$. Panels (g)-(l) refer to the trivial to topological phase transition. The chemical potential is chosen to be: (g) $4.0$, (h) $3.999$, (i) $3.995$, (j) $3.99$, (k)  $3.95$, and (l) $3.9$. }
\label{QPT_CB}
\end{figure*}

We examine the EE for NCBs for varying system sizes $M = 50 \times 50, 60 \times 60, 70 \times 70, 80 \times 80$ [see Fig. \ref{NCB_EE&ES}(a)].  The subsystem dimension is chosen to be $M_\Omega = \left(  40 \times N_y \right) $, where $N_y = N_x$ are varied as per the dimension given in the Fig. \ref{NCB_EE&ES} (a). As a function of the chemical potential $\mu$, the entropy $S_\Omega(\mu)$ displays pronounced non-monotonic variation. Peaks and dips in the EE correspond to TQPTs, including changes in the nodal connectivity of the energy spectrum, discussed in the Appendix \ref{app_defects}. Compared to the contractible case, the EE for the same $\mu$ and $N$ is significantly higher, reflecting non-local entanglement across the non-contractible cut. We restrict the dimension of $M$, since in an NCB, the EE follows the subsystem size, which is evident in the Fig. \ref{NCB_EE&ES}(a). This enhanced entropy can include contributions from several sources: (i) coupling to global topological degrees of freedom, (ii) long-range correlations crossing the non-contractible boundary, and (iii) cumulative effects of ground-state degenerate modes encircling the system.

Given the circumstances, the entanglement spectrum (ES) offers a more detailed window into these phenomena. For NCBs, the ES may reveal features inaccessible to EE alone, such as low-lying degenerate modes and topological sector decomposition. Figure \ref{NCB_EE&ES}(b) shows the ES for an NCB across various values of $\mu$. Near the transition point at $\mu \approx 0$, a notable accumulation of maximally entangled modes around $\xi = 0$ is observed, with evident dispersion of bulk modes away from $\xi=0$. As we move away from $\mu=0$, these modes become more densely packed, forming a flat band structure reminiscent of degenerate modes in gapped topological systems. In our case, one can interpret them as the signature of the presence of Majorana zero modes. This spectral flattening reflects long-range correlations across the NCB boundary. They arise from interference between nodal quasiparticles traversing through the topological sector. Remarkably, these entanglement features are topologically protected, despite the absence of a full energy gap. Their robust structure and degenerate nature exhibit strong qualitative parallels with Majorana zero modes and the edge spectra anticipated by the Li-Haldane conjecture    \cite{Li_Haldane_2008}. The persistence and symmetry of the low-lying (or the maximally entangled) entanglement modes across system sizes underscore the entanglement spectrum’s role as a robust and reliable diagnostic of non-local entanglement behavior even in gapless phases.

\subsection{Entanglement response near TQPTs}
\label{ES to QPT}

After establishing that entanglement measures such as EE and ES, when studied for CB and NCB partitions, are useful to capture the TQPT even in gapless systems, we focus on discussing the system near the QCPs that cause TQPTs. For this purpose, we compute the derivative of the entanglement entropy (EE) with respect to the chemical potential, denoted as $\left(\frac{dS_\Omega}{d\mu}\right)$, which provides valuable insights into the sensitivity of entanglement to variations in the chemical potential $\mu$. This quantity reveals how quantum correlations are redistributed as the system's chemical potential is tuned, essential for understanding TQPTs and QCPs that cause them. The derivative of EE with respect to $\mu$ exhibits more complicated, non-monotonic behavior, often showing sharp features or singularities at critical points. The response is more sensitive to low-energy excitations, leading to divergent behavior near TQPTs. The Fig. \ref{dEE}, reveals two major QCPs: (i) $\abs{\mu} \to 0$ ; (ii) $\abs{\mu} \to 4$. The opposite signs of the derivative of the entanglement entropy with respect to $\mu$ on either side of the transition indicate an asymmetric entanglement response across the QCPs. In Appendix \ref{app_TI}, we discuss the topological aspect of this asymmetry to enhance the understanding of the TQPT in the gapless regime. Notably, the derivative of the entanglement entropy (EE) with respect to the chemical potential $\mu$ in the NCB system exhibits dominant peaks compared to that in the CB system. This again confirms the robustness of entanglement measures in NCB systems, but the overall results remain broadly similar in both cases when EE is studied. Having located the QCPs, we examine the behavior of the ES in their vicinity.

Figures \ref{QPT_CB}(a)-(f) depict the ES within the topological gapless phase of the model near the QCP-I, $\abs{\mu} \to 0$. Interestingly, although the system is known to host topological order in this parameter regime, the ES derived from the CB bipartition does not clearly signal any TQPT. This lack of a distinctive spectral signature indicates that the CB may not be sensitive enough to detect changes in the topological character of gapless phases, particularly when bulk excitations are present and the entanglement structure is more subtle. However, a transition from a trivial gapped phase to a topological gapless phase near the QCP-II $\abs{\mu} \to 4$ is clearly manifested in the ES for the CB partition, as illustrated in Fig. \ref{QPT_CB}(g)-(l). Here, the emergence of low-lying entanglement levels resembling edge states serves as a hallmark of nontrivial topology. Despite this, the CB partition fails to provide a complete picture of the topological structure, especially in gapless regimes where conventional indicators become less reliable. These findings highlight a key limitation of the CB approach: detecting topological transitions only due to local parameters. It is insensitive primarily to subtle topological signatures due to non-local order parameters.
\begin{figure*}
\includegraphics[width=1\textwidth]{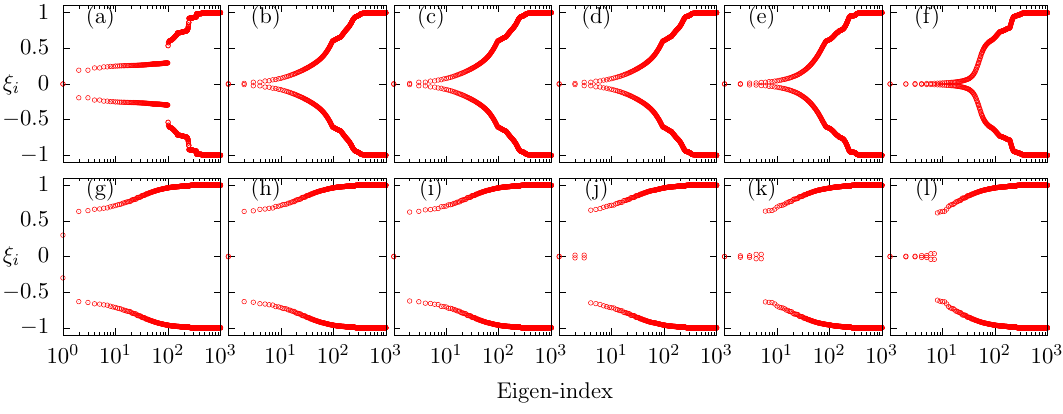}
\caption{The ES for an NCB subsystem near the QCPs. Panels (a)-(f) refer to the TQPT happening around $\mu=0$, where  $\abs{\mu}>0$ is a topological phase. The chemical potential is chosen to be: (a) $0.0$, (b) $0.001$, (c) $0.005$, (d) $0.01$, (e)  $0.05$, and (f) $0.1$. Panels (g)-(l) refer to the trivial to topological phase transition. The chemical potential is chosen to be: (g) $4.0$, (h) $3.999$, (i) $3.995$, (j) $3.99$, (k)  $3.95$, and (l) $3.9$. }
\label{QPT_NCB}
\end{figure*}

In contrast, the NCB case, which involves more intricate global sector divisions, proves to be far more effective in unveiling the intrinsic topological features of the model. As shown in Fig. \ref{QPT_NCB}(a)-(f), the ES obtained from the NCB partition near the QCP-I exhibits clear evidence of a TQPT even within the gapless topological regime. Although the system remains gapless in this region, the ES shows significant restructuring, indicating a qualitative change in the entanglement structure consistent with a TQPT.

Moreover, the transition from a trivial to a topological phase is far more sharply captured using the NCB partition (near QCP-II) in the Fig. \ref{QPT_NCB}(g)-(l). The low-lying entanglement levels (or maximally entangled level) reveal a clear separation from the higher bulk entanglement bands, forming a well-defined entanglement gap, a signature often associated with topological order. This stark contrast with the CB case underlines the importance of choosing an appropriate subsystem geometry when probing topological properties using entanglement measures.

These results collectively emphasize that while conventional partitions might suffice for detecting TQPTs caused by local order parameters in gapped systems, more sophisticated bipartitions such as the NCB are essential for revealing the complete details about the topological order, especially in systems with gapless excitations or non-local entanglement. The enhanced sensitivity of the NCB partition makes it a powerful tool for characterizing topological order in complex systems, where conventional order parameters fail or are ill-defined.

\section{Discussion and experimental realisations}
\label{Conclusion}

In this work, we have explored the topological properties of a 2D gapless $p$-wave SC using various entanglement measures as a diagnostic tool. Our analysis focused on three key entanglement-based measures: the entanglement spectrum (ES), the entanglement entropy (EE), and the derivative of the entanglement entropy with respect to chemical potential $\left( \frac{dS_\Omega}{d\mu} \right)$, under contractible and non-contractible bipartitions. We find that the behavior of entanglement significantly differs between contractible bipartitions (CBs) and non-contractible bipartitions (NCBs), revealing a rich interplay between topology, quantum correlations, and criticality in gapless BdG systems. 

The EE exhibits area-law scaling for CB subsystems that remains independent of subsystem size, indicating the domination of short-range entanglement near the boundary. Nonetheless, EE remains sensitive to topological quantum phase transitions (TQPTs), as evident from non-monotonic variations and entropy peaks as the chemical potential $\mu$ is tuned across gap-closing points. The ES complements this picture by capturing qualitative changes in boundary mode structures, signaling shifts in the system's topological character.

In contrast, NCB subsystems probe global aspects of the wavefunction and reveal long-range entanglement. Here, the EE grows with subsystem size, reflecting the presence of extended quantum correlations. The ES under NCBs exhibits notable features: a dense accumulation of low-lying entanglement modes near $\xi \to 0$, and a nearly flat band structure that becomes sharper and more pronounced with increasing system size. These features are topologically protected and arise due to interference effects among nodal quasiparticles in the nontrivial topological sector.

Importantly, our results demonstrate that the derivative of EE with respect to $\mu$ is an effective diagnostic for identifying TQPTs, with clear peaks marking quantum critical points (QCPs). Furthermore, near these QCPs, entanglement measures confirm additional signatures of topological behavior, including the topological defects and quantized invariants. The appearance of topological defects gives rise to protected edge states, and quantized invariants signal a TQPT. Entanglement measures capture these topological behaviors, providing further evidence of the system's nontrivial topology. This study underscores the power of entanglement-based diagnostics as a probe for unconventional topological phenomena in gapless BdG superconductors under various bipartitions.

Recent studies have increasingly focused on realizing 2D BdG systems exhibiting topological superconductivity, including the hybrid structures combining spin-spiral magnetic textures with conventional $s$-wave superconductors, which have demonstrated signatures consistent with emergent $(p_x + p_y)$-type pairing and gapless topological phases \cite{exp_realisation_Arijit}. Recent advances in `topo-electronics' provide additional proof for realising topological models    \cite{top_circuits}. These platforms offer promising routes to engineer and probe 2D topological superconducting states similar to our model's. Also, entanglement measures such as entanglement entropy have been experimentally accessed in various quantum simulators, including photonic lattices, trapped ions, and superconducting qubit arrays    \cite{EE_optical_lattice, EE_phononic_system}. These experiments have confirmed that entanglement metrics serve as reliable diagnostics of phase transitions and many-body correlations beyond the specifics of any single physical model. Together, these advances highlight the possibility that our proposed 2D model in the BdG framework can be a future platform to explore and unify spectral and entanglement-based signatures of topological phases in complex quantum materials.

\appendix % Start of appendices

\section{Additional information about the topology characteristics of the model near the QCPs present in the model}
\label{app}

The model under consideration preserves all three fundamental symmetries: chiral, time-reversal, and particle-hole \cite{2DKitaev, zhang2019, Srinidhi_25}. The TRS operator $\mathcal{T}$ for spinless systems is,  simply the conjugation operator $ \mathcal{K}$, whichgives, $\mathcal{T}^2=+\mathcal{I}$. This Appendix briefly presents the topological characteristics of the $p$-wave SC model. More specifically, we focus on the regime described in Sec. \ref{ES to QPT}, and depict how the topology of the model plays a critical role in understanding the entanglement signatures of the model.

\subsection{Topological Invariant}
\label{app_TI}

\begin{figure}
\includegraphics[width=1\columnwidth]{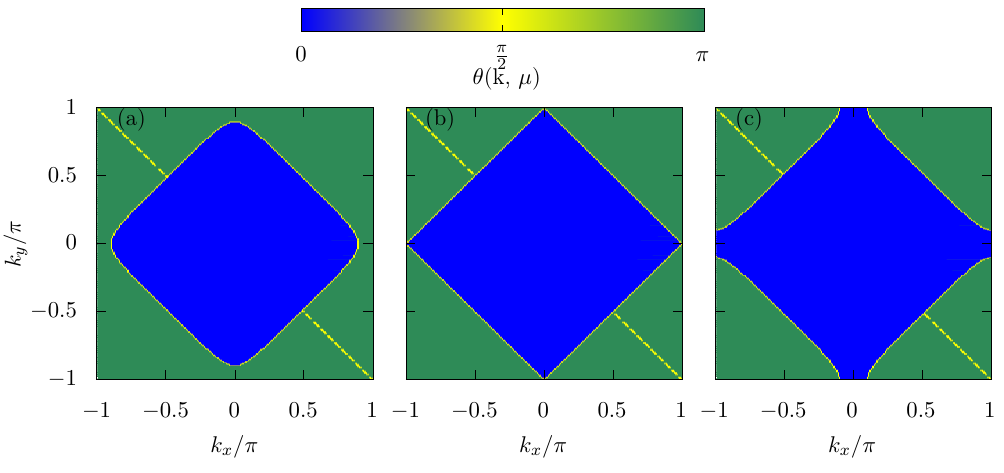}
\caption{Figure corresponds to the topological invariant of the model Hamiltonian. The chemical potential is varied as: (a) $\mu = 0.1$, (b) $\mu =0.0$, and (c) $\mu = - 0.1$. The other parameters remain similar to previous results.}
\label{Z2_fig}
\end{figure}

To identify the two topological phases that undergo a TQPT  at the QCP-I, we present a topological invariant that characterizes the topological phase of the model. Systems in the DIII topological class are characterized by a $\mathbb{Z}_2$ topological invariant, which distinguishes between two distinct topological phases. For inversion-symmetric systems, the $\mathbb{Z}_2$ invariant can be extracted using the Pfaffian of a matrix constructed from the time-reversal operator acting on the occupied states. Specifically, the matrix elements are given by   \cite{Chen_2016, WChen_2016, Srinidhi_25}
\be
m_{\alpha\beta}(\boldsymbol{k},\mu) = \langle\psi_\alpha(\boldsymbol{k}, \mu)|\mathcal{T} |\psi_\beta(\boldsymbol{k}, \mu)\rangle,
\label{theta}
\ee
where $\ket{\psi_\alpha(\boldsymbol{k}, \mu)}$ is the $\alpha$-th occupied eigenstate at momentum $\boldsymbol{k}$, and $\mu$ denotes a tunable parameter in the system, representing the chemical potential of the nodal quasiparticle in the system. The $\mathbb{Z}_2$ invariant is computed from the argument of the Pfaffian of the matrix $m$, as follows:
\be
\theta(\boldsymbol{k}, \mu) = \arg \left[\mathrm{Pf}(m_{\alpha\beta})\right],
\ee
where, $\arg$ defines the argument of the complex number $\mathrm{Pf}(m)$. Considering the Hamiltonian in Eq. \eqref{hk}, the eigenstates take the form:
\be
\psi_\pm(\boldsymbol{k}) = \begin{pmatrix} E(\boldsymbol{k}) \pm \epsilon(\boldsymbol{k}) \\ -i \chi(\boldsymbol{k}) \end{pmatrix},
\ee
where, the corresponding eigenenergies are $ E(\boldsymbol{k}) = \pm \sqrt{\epsilon(\boldsymbol{k})^2 + \chi(\boldsymbol{k})^2}$. Figure \ref{Z2_fig} depicts the evolution of $\theta(\boldsymbol{k}, \mu)$ for different values of the chemical potential $\mu$. These specific values highlight the distinction between the two topological phases in the gapless regime, which undergoes a topological quantum phase transition (TQPT). As $\mu$ varies, the system transitions between different topological phases, which is reflected in the structure of $\theta(\boldsymbol{k}, \mu)$. They signal the emergence of momentum-space vortices associated with non-trivial topological behavior. Although these vortices do not correspond to physical vortex cores in real space, the local behavior of $\theta(\boldsymbol{k}, \mu)$ near such features provides insight into the presence of topological defects with opposite winding numbers in the Brillouin zone. As expected, the net topological charge associated with this $\mathbb{Z}_2$ invariant remains zero, yet it clearly distinguishes between the two topologically distinct phases in the gapless regime. Next, the presence of vortices has been discussed in detail.

\subsection{ Topological Defects}
\label{app_defects}

We revisit supporting analysis from earlier studies relevant to our current work   \cite{zhang2019, Srinidhi_25}. Band touching occurs when the components of the Hamiltonian vector vanish simultaneously. By interpreting these components as forming a vector field, one can identify vortex-like structures within the field   \cite{Hasan2021-qz,Jia2016-qo}. These vortices correspond to two-dimensional topological defects such as Dirac cones or Weyl nodes. Despite their structural similarity to topologically nontrivial features, these defects do not carry a net topological charge, as they arise from overlapping Weyl nodes of opposite chirality whose contributions cancel   \cite{Funfhaus_2022}. This cancellation reflects the monopole-like character of Weyl points in momentum space, which act as sources or sinks of Berry curvature. Such defects are topologically protected, remain stable in the presence of disorder, and can only be removed via a gap closing and reopening induced by variations in system parameters. To identify analogous behavior in our model, we analyze the Hamiltonian components and detect the formation of new singularities near the degenerate points located at $ (k_{x_0}, k_{y_0}) $ in the Brillouin zone. These singularities satisfy the following conditions:

\begin{align}
& \Delta [ \sin(k_{x_0}) + \sin(k_{y_0}) ] = 0 \label{eq:del} \\
& \mu + 2w [ \cos(k_{x_0}) + \cos(k_{y_0}) ] = 0 \label{eq:w}
\end{align}

Under the parameter regions indicated in Fig. (\ref{QPT_CB})–(\ref{QPT_NCB}), we analyze the topological defects associated with various TQPTs. The solutions to Eqs. \eqref{eq:del} and \eqref{eq:w} define the loci of degenerate points in the Brillouin zone (BZ), which may appear as either isolated points (point defects) or extended lines (line defects), depending on the system parameters. Below, we analyze two representative cases for $\mu $, with $ w = 1 $, $\Delta = 0.8 $, and BZ defined as $ k_x, k_y \in [0, 2\pi] $.

%\underline{\textbf{Case-1 $\mu = 4$ (Point Defect):}} \quad\\

\begin{itemize}

\item \textbf{Case 1: $\mu = 4$ (Point Defect):}

From Eq. \eqref{eq:del}: $ \sin(k_{y_0}) = -\sin(k_{x_0})$. This condition is satisfied when $ k_{y_0} = 2\pi - k_{x_0} $ within the BZ. Substituting this into Eq. \eqref{eq:w}:
\be
4 + 2 \left[ \cos(k_{x_0}) + \cos(2\pi - k_{x_0}) \right] = -2
\ee
Since $\cos(2\pi - k_{x_0}) = \cos(k_{x_0}) $, we obtain:
\be
2\cos(k_{x_0}) = -2   \Rightarrow \boxed{ k_{x_0, y_0} =(\pi, \pi)} 
\label{P1}
\ee
Thus, the band touching point in Eq. \eqref{P1} is an isolated point defect in the Brillouin zone, representing a local singularity in momentum space.

%\underline{\textbf{Case 2 -$\mu = 0 $ (Line Defect):}} \quad\\

\item \textbf{Case 2: $\mu = 0$ (Line Defect):}

In this case, Eq. \eqref{eq:w} becomes:
\be
\begin{split}
\cos(k_{y_0}) &= -\cos(k_{x_0}) \\
\sin(k_{y_0}) &= -\sin(k_{x_0})
\end{split}
\ee
Together, these imply:\quad $\boxed{ k_{y_0} = \pi + k_{x_0} }$.
\end{itemize}
This defines a continuous line in the Brillouin zone along which the band gap vanishes. Therefore, at $\mu = 0$, the system exhibits a line defect, or a nodal line.

\section{Additional Information about the symmetry class of the $p$-wave SC}
\label{app_symm}

 For spin$-\frac{1}{2}$ cases, the Hamiltonian in the Eq. \eqref{H_kspace} takes the form,
\begin{subequations}
\be
H = \sum_{\mathbf{k}, \mathbf{s}} \Psi^\dagger_{\mathbf{k}, \mathbf{s}} \, \mathcal{H}(\mathbf{k}, \mathbf{s}) \, \Psi_{\mathbf{k}, \mathbf{s}}, 
\ee
where the Nambu spinor and the $\mathcal{H}(\mathbf{k})$ are given as,
\be
\Psi^\dagger_{\mathbf{k}, \mathbf{s}} = \begin{pmatrix}
c^\dagger_{\mathbf{k}, \uparrow} &
c_{-\mathbf{k}, \uparrow} &
c^\dagger_{\mathbf{k}, \downarrow} &
c_{-\mathbf{k}, \downarrow}
\end{pmatrix}.
\ee
\be
\mathcal{H}(\mathbf{k}, \mathbf{s}) = 
\left[ \epsilon(\mathbf{k}) \tau_z 
+ \chi(\mathbf{k})\tau_y \right] \otimes \sigma_0,
\ee
\end{subequations}
Here, the Pauli matrices $\tau_i$ act in the quasiparticle (Nambu) basis, while $\sigma_0$ is the identity matrix in spin space. The time-reversal operator `$\mathbbm{T}$' for spin-$ \frac{1}{2} $ fermions in Nambu space is,
\be
\mathbbm{T} =(  \mathcal{T} \otimes i \sigma_y )  \mathcal{K},
\qquad \Rightarrow \qquad
\mathbbm{T}^2 =- \mathcal{I}
\ee
where $\tau_0$ is the identity matrix in the Nambu basis and $\sigma$ denotes the spin space. To check TRS, we act with $ \mathbbm{T} $:
\begin{align}
\mathbbm{T} \mathcal{H}(\mathbf{k}) \mathbbm{T}^{-1} 
&= \left[ \tau_0 \otimes i \sigma_y \right] \left[ \mathcal{K}  \mathcal{H}(\mathbf{k})  \mathcal{K} \right] \left[ \tau_0 \otimes -i \sigma_y \right] \notag \\
&= \mathcal{H}(-\mathbf{k})
\end{align}
This follows from:
\be
\epsilon(\mathbf{k}) = \epsilon(-\mathbf{k}), 
\qquad 
\chi(-\mathbf{k}) = -\chi(\mathbf{k})
\ee
Thus, $ \mathcal{H}(\mathbf{k}) $ transforms correctly under time reversal:
\be
\mathbbm{T} \mathcal{H}(\mathbf{k}) \mathbbm{T}^{-1} = \mathcal{H}(-\mathbf{k})
\ee
Since $ \mathbbm{T}^2 = -\mathcal{I} $, by Kramers' theorem, every energy eigenstate is at least doubly degenerate. This holds throughout the 2D BZ. In particular, the TRS operator satisfies $\mathbbm{T}^2 = -\mathcal{I}$, indicating that the system belongs to the symmetry class DIII, according to the Altland-Zirnbauer classification and the periodic table of topological insulators and superconductors \cite{symmetry_class1, symmetry_class2, symmetry_class3}. In this article, we analyze the spinless (or single-spin polarised) version of the model. Since the TRITOPS decouples into two copies of identical spin sectors in the limit of vanishing interaction term between spin and fermionic degree of freedom, focusing on one spin sector retains the complete physics \cite{TRI_review}. Importantly, this reduced model still exhibits nontrivial gapless phases, characterized by nodal points in the Brillouin zone arising from the interplay of pairing and kinetic terms. This is similar to quantum Hall systems, where each spin component can have a different topological number. Still, the whole system may be time-reversal invariant (if opposite spin topological numbers cancel) \cite{Srinidhi_25}. The topological features and entanglement structure remain representative of the whole system.

\bibliography{Ref} 

%apsrev4-2.bst 2019-01-14 (MD) hand-edited version of apsrev4-1.bst
%Control: key (0)
%Control: author (8) initials jnrlst
%Control: editor formatted (1) identically to author
%Control: production of article title (0) allowed
%Control: page (0) single
%Control: year (1) truncated
%Control: production of eprint (0) enabled
\begin{thebibliography}{72}%
\makeatletter
\providecommand \@ifxundefined [1]{%
 \@ifx{#1\undefined}
}%
\providecommand \@ifnum [1]{%
 \ifnum #1\expandafter \@firstoftwo
 \else \expandafter \@secondoftwo
 \fi
}%
\providecommand \@ifx [1]{%
 \ifx #1\expandafter \@firstoftwo
 \else \expandafter \@secondoftwo
 \fi
}%
\providecommand \natexlab [1]{#1}%
\providecommand \enquote  [1]{``#1''}%
\providecommand \bibnamefont  [1]{#1}%
\providecommand \bibfnamefont [1]{#1}%
\providecommand \citenamefont [1]{#1}%
\providecommand \href@noop [0]{\@secondoftwo}%
\providecommand \href [0]{\begingroup \@sanitize@url \@href}%
\providecommand \@href[1]{\@@startlink{#1}\@@href}%
\providecommand \@@href[1]{\endgroup#1\@@endlink}%
\providecommand \@sanitize@url [0]{\catcode `\\12\catcode `\$12\catcode
  `\&12\catcode `\#12\catcode `\^12\catcode `\_12\catcode `\%12\relax}%
\providecommand \@@startlink[1]{}%
\providecommand \@@endlink[0]{}%
\providecommand \url  [0]{\begingroup\@sanitize@url \@url }%
\providecommand \@url [1]{\endgroup\@href {#1}{\urlprefix }}%
\providecommand \urlprefix  [0]{URL }%
\providecommand \Eprint [0]{\href }%
\providecommand \doibase [0]{https://doi.org/}%
\providecommand \selectlanguage [0]{\@gobble}%
\providecommand \bibinfo  [0]{\@secondoftwo}%
\providecommand \bibfield  [0]{\@secondoftwo}%
\providecommand \translation [1]{[#1]}%
\providecommand \BibitemOpen [0]{}%
\providecommand \bibitemStop [0]{}%
\providecommand \bibitemNoStop [0]{.\EOS\space}%
\providecommand \EOS [0]{\spacefactor3000\relax}%
\providecommand \BibitemShut  [1]{\csname bibitem#1\endcsname}%
\let\auto@bib@innerbib\@empty
%</preamble>
\bibitem [{\citenamefont {Qi}\ and\ \citenamefont
  {Zhang}(2011)}]{TI_and_TSC_2011}%
  \BibitemOpen
  \bibfield  {author} {\bibinfo {author} {\bibfnamefont {X.-L.}\ \bibnamefont
  {Qi}}\ and\ \bibinfo {author} {\bibfnamefont {S.-C.}\ \bibnamefont {Zhang}},\
  }\bibfield  {title} {\bibinfo {title} {Topological insulators and
  superconductors},\ }\href {https://doi.org/10.1103/RevModPhys.83.1057}
  {\bibfield  {journal} {\bibinfo  {journal} {Rev. Mod. Phys.}\ }\textbf
  {\bibinfo {volume} {83}},\ \bibinfo {pages} {1057} (\bibinfo {year}
  {2011})}\BibitemShut {NoStop}%
\bibitem [{\citenamefont {Wen}(2017)}]{Zoo_TQPT}%
  \BibitemOpen
  \bibfield  {author} {\bibinfo {author} {\bibfnamefont {X.-G.}\ \bibnamefont
  {Wen}},\ }\bibfield  {title} {\bibinfo {title} {Colloquium: Zoo of
  quantum-topological phases of matter},\ }\href
  {https://doi.org/10.1103/RevModPhys.89.041004} {\bibfield  {journal}
  {\bibinfo  {journal} {Rev. Mod. Phys.}\ }\textbf {\bibinfo {volume} {89}},\
  \bibinfo {pages} {041004} (\bibinfo {year} {2017})}\BibitemShut {NoStop}%
\bibitem [{\citenamefont {Scaffidi}\ \emph {et~al.}(2017)\citenamefont
  {Scaffidi}, \citenamefont {Parker},\ and\ \citenamefont
  {Vasseur}}]{GSPT_2017}%
  \BibitemOpen
  \bibfield  {author} {\bibinfo {author} {\bibfnamefont {T.}~\bibnamefont
  {Scaffidi}}, \bibinfo {author} {\bibfnamefont {D.~E.}\ \bibnamefont
  {Parker}},\ and\ \bibinfo {author} {\bibfnamefont {R.}~\bibnamefont
  {Vasseur}},\ }\bibfield  {title} {\bibinfo {title} {Gapless
  symmetry-protected topological order},\ }\href
  {https://doi.org/10.1103/PhysRevX.7.041048} {\bibfield  {journal} {\bibinfo
  {journal} {Phys. Rev. X}\ }\textbf {\bibinfo {volume} {7}},\ \bibinfo {pages}
  {041048} (\bibinfo {year} {2017})}\BibitemShut {NoStop}%
\bibitem [{\citenamefont {Thorngren}\ \emph {et~al.}(2021)\citenamefont
  {Thorngren}, \citenamefont {Vishwanath},\ and\ \citenamefont
  {Verresen}}]{Intrinsic_GPT}%
  \BibitemOpen
  \bibfield  {author} {\bibinfo {author} {\bibfnamefont {R.}~\bibnamefont
  {Thorngren}}, \bibinfo {author} {\bibfnamefont {A.}~\bibnamefont
  {Vishwanath}},\ and\ \bibinfo {author} {\bibfnamefont {R.}~\bibnamefont
  {Verresen}},\ }\bibfield  {title} {\bibinfo {title} {Intrinsically gapless
  topological phases},\ }\href {https://doi.org/10.1103/PhysRevB.104.075132}
  {\bibfield  {journal} {\bibinfo  {journal} {Phys. Rev. B}\ }\textbf {\bibinfo
  {volume} {104}},\ \bibinfo {pages} {075132} (\bibinfo {year}
  {2021})}\BibitemShut {NoStop}%
\bibitem [{\citenamefont {Zhou}\ \emph {et~al.}(2025)\citenamefont {Zhou},
  \citenamefont {Wang},\ and\ \citenamefont {Pan}}]{Floquet_GPT}%
  \BibitemOpen
  \bibfield  {author} {\bibinfo {author} {\bibfnamefont {L.}~\bibnamefont
  {Zhou}}, \bibinfo {author} {\bibfnamefont {R.}~\bibnamefont {Wang}},\ and\
  \bibinfo {author} {\bibfnamefont {J.}~\bibnamefont {Pan}},\ }\bibfield
  {title} {\bibinfo {title} {Gapless higher-order topology and corner states in
  floquet systems},\ }\href {https://doi.org/10.1103/PhysRevResearch.7.023079}
  {\bibfield  {journal} {\bibinfo  {journal} {Phys. Rev. Res.}\ }\textbf
  {\bibinfo {volume} {7}},\ \bibinfo {pages} {023079} (\bibinfo {year}
  {2025})}\BibitemShut {NoStop}%
\bibitem [{\citenamefont {Geier}\ \emph {et~al.}(2018)\citenamefont {Geier},
  \citenamefont {Trifunovic}, \citenamefont {Hoskam},\ and\ \citenamefont
  {Brouwer}}]{II_order_TI_SC}%
  \BibitemOpen
  \bibfield  {author} {\bibinfo {author} {\bibfnamefont {M.}~\bibnamefont
  {Geier}}, \bibinfo {author} {\bibfnamefont {L.}~\bibnamefont {Trifunovic}},
  \bibinfo {author} {\bibfnamefont {M.}~\bibnamefont {Hoskam}},\ and\ \bibinfo
  {author} {\bibfnamefont {P.~W.}\ \bibnamefont {Brouwer}},\ }\bibfield
  {title} {\bibinfo {title} {Second-order topological insulators and
  superconductors with an order-two crystalline symmetry},\ }\href
  {https://doi.org/10.1103/PhysRevB.97.205135} {\bibfield  {journal} {\bibinfo
  {journal} {Phys. Rev. B}\ }\textbf {\bibinfo {volume} {97}},\ \bibinfo
  {pages} {205135} (\bibinfo {year} {2018})}\BibitemShut {NoStop}%
\bibitem [{\citenamefont {Berg}\ \emph {et~al.}(2008)\citenamefont {Berg},
  \citenamefont {Chen},\ and\ \citenamefont {Kivelson}}]{Nodal_Sc}%
  \BibitemOpen
  \bibfield  {author} {\bibinfo {author} {\bibfnamefont {E.}~\bibnamefont
  {Berg}}, \bibinfo {author} {\bibfnamefont {C.-C.}\ \bibnamefont {Chen}},\
  and\ \bibinfo {author} {\bibfnamefont {S.~A.}\ \bibnamefont {Kivelson}},\
  }\bibfield  {title} {\bibinfo {title} {Stability of nodal quasiparticles in
  superconductors with coexisting orders},\ }\href
  {https://doi.org/10.1103/PhysRevLett.100.027003} {\bibfield  {journal}
  {\bibinfo  {journal} {Phys. Rev. Lett.}\ }\textbf {\bibinfo {volume} {100}},\
  \bibinfo {pages} {027003} (\bibinfo {year} {2008})}\BibitemShut {NoStop}%
\bibitem [{\citenamefont {Amin}\ and\ \citenamefont
  {Smirnov}(2004)}]{Nodal_Qubit}%
  \BibitemOpen
  \bibfield  {author} {\bibinfo {author} {\bibfnamefont {M.~H.~S.}\
  \bibnamefont {Amin}}\ and\ \bibinfo {author} {\bibfnamefont {A.~Y.}\
  \bibnamefont {Smirnov}},\ }\bibfield  {title} {\bibinfo {title}
  {Quasiparticle decoherence in $d$-wave superconducting qubits},\ }\href
  {https://doi.org/10.1103/PhysRevLett.92.017001} {\bibfield  {journal}
  {\bibinfo  {journal} {Phys. Rev. Lett.}\ }\textbf {\bibinfo {volume} {92}},\
  \bibinfo {pages} {017001} (\bibinfo {year} {2004})}\BibitemShut {NoStop}%
\bibitem [{\citenamefont {Yu}\ \emph {et~al.}(2024)\citenamefont {Yu},
  \citenamefont {Yang}, \citenamefont {Lin},\ and\ \citenamefont
  {Jian}}]{PhysRevLett.133.026601}%
  \BibitemOpen
  \bibfield  {author} {\bibinfo {author} {\bibfnamefont {X.-J.}\ \bibnamefont
  {Yu}}, \bibinfo {author} {\bibfnamefont {S.}~\bibnamefont {Yang}}, \bibinfo
  {author} {\bibfnamefont {H.-Q.}\ \bibnamefont {Lin}},\ and\ \bibinfo {author}
  {\bibfnamefont {S.-K.}\ \bibnamefont {Jian}},\ }\bibfield  {title} {\bibinfo
  {title} {Universal entanglement spectrum in one-dimensional gapless symmetry
  protected topological states},\ }\href
  {https://doi.org/10.1103/PhysRevLett.133.026601} {\bibfield  {journal}
  {\bibinfo  {journal} {Phys. Rev. Lett.}\ }\textbf {\bibinfo {volume} {133}},\
  \bibinfo {pages} {026601} (\bibinfo {year} {2024})}\BibitemShut {NoStop}%
\bibitem [{\citenamefont {Jia}\ \emph {et~al.}(2025)\citenamefont {Jia},
  \citenamefont {Hu}, \citenamefont {Zhang}, \citenamefont {Xiao},
  \citenamefont {Wang}, \citenamefont {Wang}, \citenamefont {Ma}, \citenamefont
  {Ouyang}, \citenamefont {Zhu},\ and\ \citenamefont {Chan}}]{1DGPT_NH}%
  \BibitemOpen
  \bibfield  {author} {\bibinfo {author} {\bibfnamefont {H.}~\bibnamefont
  {Jia}}, \bibinfo {author} {\bibfnamefont {J.}~\bibnamefont {Hu}}, \bibinfo
  {author} {\bibfnamefont {R.-Y.}\ \bibnamefont {Zhang}}, \bibinfo {author}
  {\bibfnamefont {Y.}~\bibnamefont {Xiao}}, \bibinfo {author} {\bibfnamefont
  {D.}~\bibnamefont {Wang}}, \bibinfo {author} {\bibfnamefont {M.}~\bibnamefont
  {Wang}}, \bibinfo {author} {\bibfnamefont {S.}~\bibnamefont {Ma}}, \bibinfo
  {author} {\bibfnamefont {X.}~\bibnamefont {Ouyang}}, \bibinfo {author}
  {\bibfnamefont {Y.}~\bibnamefont {Zhu}},\ and\ \bibinfo {author}
  {\bibfnamefont {C.~T.}\ \bibnamefont {Chan}},\ }\bibfield  {title} {\bibinfo
  {title} {Unconventional topological edge states in one-dimensional
  non-{H}ermitian gapless systems stemming from nonisolated hypersurface
  singularities},\ }\href {https://doi.org/10.1103/PhysRevLett.134.206603}
  {\bibfield  {journal} {\bibinfo  {journal} {Phys. Rev. Lett.}\ }\textbf
  {\bibinfo {volume} {134}},\ \bibinfo {pages} {206603} (\bibinfo {year}
  {2025})}\BibitemShut {NoStop}%
\bibitem [{\citenamefont {Benalcazar}\ and\ \citenamefont
  {Cerjan}(2020)}]{TI_GPT}%
  \BibitemOpen
  \bibfield  {author} {\bibinfo {author} {\bibfnamefont {W.~A.}\ \bibnamefont
  {Benalcazar}}\ and\ \bibinfo {author} {\bibfnamefont {A.}~\bibnamefont
  {Cerjan}},\ }\bibfield  {title} {\bibinfo {title} {Bound states in the
  continuum of higher-order topological insulators},\ }\href
  {https://doi.org/10.1103/PhysRevB.101.161116} {\bibfield  {journal} {\bibinfo
   {journal} {Phys. Rev. B}\ }\textbf {\bibinfo {volume} {101}},\ \bibinfo
  {pages} {161116} (\bibinfo {year} {2020})}\BibitemShut {NoStop}%
\bibitem [{\citenamefont {Wang}\ \emph {et~al.}(2017)\citenamefont {Wang},
  \citenamefont {Lin}, \citenamefont {Zhang},\ and\ \citenamefont
  {Song}}]{2DKitaev}%
  \BibitemOpen
  \bibfield  {author} {\bibinfo {author} {\bibfnamefont {P.}~\bibnamefont
  {Wang}}, \bibinfo {author} {\bibfnamefont {S.}~\bibnamefont {Lin}}, \bibinfo
  {author} {\bibfnamefont {G.}~\bibnamefont {Zhang}},\ and\ \bibinfo {author}
  {\bibfnamefont {Z.}~\bibnamefont {Song}},\ }\bibfield  {title} {\bibinfo
  {title} {Topological gapless phase in {K}itaev model on square lattice},\
  }\bibfield  {journal} {\bibinfo  {journal} {Scientific Reports}\ }\textbf
  {\bibinfo {volume} {7}},\ \href {https://doi.org/10.1038/s41598-017-17334-w}
  {10.1038/s41598-017-17334-w} (\bibinfo {year} {2017})\BibitemShut {NoStop}%
\bibitem [{\citenamefont {Zhang}\ \emph {et~al.}(2019)\citenamefont {Zhang},
  \citenamefont {Wang},\ and\ \citenamefont {Song}}]{zhang2019}%
  \BibitemOpen
  \bibfield  {author} {\bibinfo {author} {\bibfnamefont {K.~L.}\ \bibnamefont
  {Zhang}}, \bibinfo {author} {\bibfnamefont {P.}~\bibnamefont {Wang}},\ and\
  \bibinfo {author} {\bibfnamefont {Z.}~\bibnamefont {Song}},\ }\bibfield
  {title} {\bibinfo {title} {{M}ajorana flat band edge modes of topological
  gapless phase in 2d {K}itaev square lattice},\ }\bibfield  {journal}
  {\bibinfo  {journal} {Scientific Reports}\ }\textbf {\bibinfo {volume} {9}},\
  \href {https://doi.org/10.1038/s41598-019-41529-y}
  {10.1038/s41598-019-41529-y} (\bibinfo {year} {2019})\BibitemShut {NoStop}%
\bibitem [{\citenamefont {Srinidhi}\ \emph {et~al.}(2025)\citenamefont
  {Srinidhi}, \citenamefont {Agrawal},\ and\ \citenamefont
  {Bandyopadhyay}}]{Srinidhi_25}%
  \BibitemOpen
  \bibfield  {author} {\bibinfo {author} {\bibfnamefont {S.}~\bibnamefont
  {Srinidhi}}, \bibinfo {author} {\bibfnamefont {A.}~\bibnamefont {Agrawal}},\
  and\ \bibinfo {author} {\bibfnamefont {J.~N.}\ \bibnamefont
  {Bandyopadhyay}},\ }\bibfield  {title} {\bibinfo {title} {Quasi-{M}ajorana
  modes in the p-wave {K}itaev chains on a square lattice},\ }\href
  {https://doi.org/10.1088/1361-648X/adcdae} {\bibfield  {journal} {\bibinfo
  {journal} {Journal of Physics: Condensed Matter}\ }\textbf {\bibinfo {volume}
  {37}},\ \bibinfo {pages} {205403} (\bibinfo {year} {2025})}\BibitemShut
  {NoStop}%
\bibitem [{\citenamefont {Ullah}\ \emph {et~al.}(2022)\citenamefont {Ullah},
  \citenamefont {Li}, \citenamefont {Jin}, \citenamefont {Pahari},
  \citenamefont {Yue}, \citenamefont {Xu}, \citenamefont {Balasubramanian},
  \citenamefont {Sellmyer},\ and\ \citenamefont
  {Skomski}}]{PhysRevB.106.134430}%
  \BibitemOpen
  \bibfield  {author} {\bibinfo {author} {\bibfnamefont {A.}~\bibnamefont
  {Ullah}}, \bibinfo {author} {\bibfnamefont {X.}~\bibnamefont {Li}}, \bibinfo
  {author} {\bibfnamefont {Y.}~\bibnamefont {Jin}}, \bibinfo {author}
  {\bibfnamefont {R.}~\bibnamefont {Pahari}}, \bibinfo {author} {\bibfnamefont
  {L.}~\bibnamefont {Yue}}, \bibinfo {author} {\bibfnamefont {X.}~\bibnamefont
  {Xu}}, \bibinfo {author} {\bibfnamefont {B.}~\bibnamefont {Balasubramanian}},
  \bibinfo {author} {\bibfnamefont {D.~J.}\ \bibnamefont {Sellmyer}},\ and\
  \bibinfo {author} {\bibfnamefont {R.}~\bibnamefont {Skomski}},\ }\bibfield
  {title} {\bibinfo {title} {Topological phase transitions and {B}erry-phase
  hysteresis in exchange-coupled nanomagnets},\ }\href
  {https://doi.org/10.1103/PhysRevB.106.134430} {\bibfield  {journal} {\bibinfo
   {journal} {Phys. Rev. B}\ }\textbf {\bibinfo {volume} {106}},\ \bibinfo
  {pages} {134430} (\bibinfo {year} {2022})}\BibitemShut {NoStop}%
\bibitem [{\citenamefont {Sun}\ \emph {et~al.}(2022)\citenamefont {Sun},
  \citenamefont {Huang}, \citenamefont {Wen}, \citenamefont {Li}, \citenamefont
  {Zhao}, \citenamefont {Cheng}, \citenamefont {Zhang},\ and\ \citenamefont
  {Guo}}]{PhysRevA.106.022208}%
  \BibitemOpen
  \bibfield  {author} {\bibinfo {author} {\bibfnamefont {Z.-Y.}\ \bibnamefont
  {Sun}}, \bibinfo {author} {\bibfnamefont {H.-L.}\ \bibnamefont {Huang}},
  \bibinfo {author} {\bibfnamefont {H.-X.}\ \bibnamefont {Wen}}, \bibinfo
  {author} {\bibfnamefont {M.}~\bibnamefont {Li}}, \bibinfo {author}
  {\bibfnamefont {X.}~\bibnamefont {Zhao}}, \bibinfo {author} {\bibfnamefont
  {H.-G.}\ \bibnamefont {Cheng}}, \bibinfo {author} {\bibfnamefont
  {D.}~\bibnamefont {Zhang}},\ and\ \bibinfo {author} {\bibfnamefont
  {B.}~\bibnamefont {Guo}},\ }\bibfield  {title} {\bibinfo {title}
  {Multipartite nonlocality and topological quantum phase transitions in a
  spinless fermion quantum wire with uniform and incommensurate potentials},\
  }\href {https://doi.org/10.1103/PhysRevA.106.022208} {\bibfield  {journal}
  {\bibinfo  {journal} {Phys. Rev. A}\ }\textbf {\bibinfo {volume} {106}},\
  \bibinfo {pages} {022208} (\bibinfo {year} {2022})}\BibitemShut {NoStop}%
\bibitem [{\citenamefont {Duan}\ \emph {et~al.}(2023)\citenamefont {Duan},
  \citenamefont {Wang},\ and\ \citenamefont {Chen}}]{PhysRevB.107.094415}%
  \BibitemOpen
  \bibfield  {author} {\bibinfo {author} {\bibfnamefont {L.}~\bibnamefont
  {Duan}}, \bibinfo {author} {\bibfnamefont {Y.-Z.}\ \bibnamefont {Wang}},\
  and\ \bibinfo {author} {\bibfnamefont {Q.-H.}\ \bibnamefont {Chen}},\
  }\bibfield  {title} {\bibinfo {title} {Quantum phase transitions in the
  triangular coupled-top model},\ }\href
  {https://doi.org/10.1103/PhysRevB.107.094415} {\bibfield  {journal} {\bibinfo
   {journal} {Phys. Rev. B}\ }\textbf {\bibinfo {volume} {107}},\ \bibinfo
  {pages} {094415} (\bibinfo {year} {2023})}\BibitemShut {NoStop}%
\bibitem [{\citenamefont {Iqbal}\ and\ \citenamefont
  {Schuch}(2021)}]{Beyond_TPT_1}%
  \BibitemOpen
  \bibfield  {author} {\bibinfo {author} {\bibfnamefont {M.}~\bibnamefont
  {Iqbal}}\ and\ \bibinfo {author} {\bibfnamefont {N.}~\bibnamefont {Schuch}},\
  }\bibfield  {title} {\bibinfo {title} {Entanglement order parameters and
  critical behavior for topological phase transitions and beyond},\ }\href
  {https://doi.org/10.1103/PhysRevX.11.041014} {\bibfield  {journal} {\bibinfo
  {journal} {Phys. Rev. X}\ }\textbf {\bibinfo {volume} {11}},\ \bibinfo
  {pages} {041014} (\bibinfo {year} {2021})}\BibitemShut {NoStop}%
\bibitem [{\citenamefont {Chen}\ and\ \citenamefont {Li}(2010)}]{Corr_QPT}%
  \BibitemOpen
  \bibfield  {author} {\bibinfo {author} {\bibfnamefont {Y.-X.}\ \bibnamefont
  {Chen}}\ and\ \bibinfo {author} {\bibfnamefont {S.-W.}\ \bibnamefont {Li}},\
  }\bibfield  {title} {\bibinfo {title} {Quantum correlations in topological
  quantum phase transitions},\ }\href
  {https://doi.org/10.1103/PhysRevA.81.032120} {\bibfield  {journal} {\bibinfo
  {journal} {Phys. Rev. A}\ }\textbf {\bibinfo {volume} {81}},\ \bibinfo
  {pages} {032120} (\bibinfo {year} {2010})}\BibitemShut {NoStop}%
\bibitem [{\citenamefont {Warren}\ \emph {et~al.}(2024)\citenamefont {Warren},
  \citenamefont {Sager-Smith},\ and\ \citenamefont {Mazziotti}}]{TQPT_RDM}%
  \BibitemOpen
  \bibfield  {author} {\bibinfo {author} {\bibfnamefont {S.}~\bibnamefont
  {Warren}}, \bibinfo {author} {\bibfnamefont {L.~M.}\ \bibnamefont
  {Sager-Smith}},\ and\ \bibinfo {author} {\bibfnamefont {D.~A.}\ \bibnamefont
  {Mazziotti}},\ }\bibfield  {title} {\bibinfo {title} {Topological phase
  transitions captured in the set of reduced density matrices},\ }\href
  {https://doi.org/10.1103/PhysRevB.109.045134} {\bibfield  {journal} {\bibinfo
   {journal} {Phys. Rev. B}\ }\textbf {\bibinfo {volume} {109}},\ \bibinfo
  {pages} {045134} (\bibinfo {year} {2024})}\BibitemShut {NoStop}%
\bibitem [{\citenamefont {Cirac}\ \emph {et~al.}(2021)\citenamefont {Cirac},
  \citenamefont {P\'erez-Garc\'{\i}a}, \citenamefont {Schuch},\ and\
  \citenamefont {Verstraete}}]{RevModPhys.93.045003}%
  \BibitemOpen
  \bibfield  {author} {\bibinfo {author} {\bibfnamefont {J.~I.}\ \bibnamefont
  {Cirac}}, \bibinfo {author} {\bibfnamefont {D.}~\bibnamefont
  {P\'erez-Garc\'{\i}a}}, \bibinfo {author} {\bibfnamefont {N.}~\bibnamefont
  {Schuch}},\ and\ \bibinfo {author} {\bibfnamefont {F.}~\bibnamefont
  {Verstraete}},\ }\bibfield  {title} {\bibinfo {title} {Matrix product states
  and projected entangled pair states: Concepts, symmetries, theorems},\ }\href
  {https://doi.org/10.1103/RevModPhys.93.045003} {\bibfield  {journal}
  {\bibinfo  {journal} {Rev. Mod. Phys.}\ }\textbf {\bibinfo {volume} {93}},\
  \bibinfo {pages} {045003} (\bibinfo {year} {2021})}\BibitemShut {NoStop}%
\bibitem [{\citenamefont {Che}\ \emph {et~al.}(2020)\citenamefont {Che},
  \citenamefont {Gneiting}, \citenamefont {Liu},\ and\ \citenamefont
  {Nori}}]{TQPT_ML}%
  \BibitemOpen
  \bibfield  {author} {\bibinfo {author} {\bibfnamefont {Y.}~\bibnamefont
  {Che}}, \bibinfo {author} {\bibfnamefont {C.}~\bibnamefont {Gneiting}},
  \bibinfo {author} {\bibfnamefont {T.}~\bibnamefont {Liu}},\ and\ \bibinfo
  {author} {\bibfnamefont {F.}~\bibnamefont {Nori}},\ }\bibfield  {title}
  {\bibinfo {title} {Topological quantum phase transitions retrieved through
  unsupervised machine learning},\ }\href
  {https://doi.org/10.1103/PhysRevB.102.134213} {\bibfield  {journal} {\bibinfo
   {journal} {Phys. Rev. B}\ }\textbf {\bibinfo {volume} {102}},\ \bibinfo
  {pages} {134213} (\bibinfo {year} {2020})}\BibitemShut {NoStop}%
\bibitem [{\citenamefont {Peschel}\ and\ \citenamefont
  {Eisler}(2009)}]{Peschel_2009}%
  \BibitemOpen
  \bibfield  {author} {\bibinfo {author} {\bibfnamefont {I.}~\bibnamefont
  {Peschel}}\ and\ \bibinfo {author} {\bibfnamefont {V.}~\bibnamefont
  {Eisler}},\ }\bibfield  {title} {\bibinfo {title} {Reduced density matrices
  and entanglement entropy in free lattice models},\ }\href
  {https://doi.org/10.1088/1751-8113/42/50/504003} {\bibfield  {journal}
  {\bibinfo  {journal} {Journal of Physics A: Mathematical and Theoretical}\
  }\textbf {\bibinfo {volume} {42}},\ \bibinfo {pages} {504003} (\bibinfo
  {year} {2009})}\BibitemShut {NoStop}%
\bibitem [{\citenamefont {Eisler}\ and\ \citenamefont
  {Peschel}(2014)}]{Eisler_2014}%
  \BibitemOpen
  \bibfield  {author} {\bibinfo {author} {\bibfnamefont {V.}~\bibnamefont
  {Eisler}}\ and\ \bibinfo {author} {\bibfnamefont {I.}~\bibnamefont
  {Peschel}},\ }\bibfield  {title} {\bibinfo {title} {Surface and bulk
  entanglement in free-fermion chains},\ }\href
  {https://doi.org/10.1088/1742-5468/2014/04/P04005} {\bibfield  {journal}
  {\bibinfo  {journal} {Journal of Statistical Mechanics: Theory and
  Experiment}\ }\textbf {\bibinfo {volume} {2014}},\ \bibinfo {pages} {P04005}
  (\bibinfo {year} {2014})}\BibitemShut {NoStop}%
\bibitem [{\citenamefont {Canovi}\ \emph {et~al.}(2014)\citenamefont {Canovi},
  \citenamefont {Ercolessi}, \citenamefont {Naldesi}, \citenamefont {Taddia},\
  and\ \citenamefont {Vodola}}]{PhysRevB.89.104303}%
  \BibitemOpen
  \bibfield  {author} {\bibinfo {author} {\bibfnamefont {E.}~\bibnamefont
  {Canovi}}, \bibinfo {author} {\bibfnamefont {E.}~\bibnamefont {Ercolessi}},
  \bibinfo {author} {\bibfnamefont {P.}~\bibnamefont {Naldesi}}, \bibinfo
  {author} {\bibfnamefont {L.}~\bibnamefont {Taddia}},\ and\ \bibinfo {author}
  {\bibfnamefont {D.}~\bibnamefont {Vodola}},\ }\bibfield  {title} {\bibinfo
  {title} {Dynamics of entanglement entropy and entanglement spectrum crossing
  a quantum phase transition},\ }\href
  {https://doi.org/10.1103/PhysRevB.89.104303} {\bibfield  {journal} {\bibinfo
  {journal} {Phys. Rev. B}\ }\textbf {\bibinfo {volume} {89}},\ \bibinfo
  {pages} {104303} (\bibinfo {year} {2014})}\BibitemShut {NoStop}%
\bibitem [{\citenamefont {Eisler}\ \emph {et~al.}(2020)\citenamefont {Eisler},
  \citenamefont {Di~Giulio}, \citenamefont {Tonni},\ and\ \citenamefont
  {Peschel}}]{Eisler_2020}%
  \BibitemOpen
  \bibfield  {author} {\bibinfo {author} {\bibfnamefont {V.}~\bibnamefont
  {Eisler}}, \bibinfo {author} {\bibfnamefont {G.}~\bibnamefont {Di~Giulio}},
  \bibinfo {author} {\bibfnamefont {E.}~\bibnamefont {Tonni}},\ and\ \bibinfo
  {author} {\bibfnamefont {I.}~\bibnamefont {Peschel}},\ }\bibfield  {title}
  {\bibinfo {title} {Entanglement hamiltonians for non-critical quantum
  chains},\ }\href {https://doi.org/10.1088/1742-5468/abb4da} {\bibfield
  {journal} {\bibinfo  {journal} {Journal of Statistical Mechanics: Theory and
  Experiment}\ }\textbf {\bibinfo {volume} {2020}},\ \bibinfo {pages} {103102}
  (\bibinfo {year} {2020})}\BibitemShut {NoStop}%
\bibitem [{\citenamefont {Eisler}\ \emph {et~al.}(2022)\citenamefont {Eisler},
  \citenamefont {Tonni},\ and\ \citenamefont {Peschel}}]{Eisler_2022}%
  \BibitemOpen
  \bibfield  {author} {\bibinfo {author} {\bibfnamefont {V.}~\bibnamefont
  {Eisler}}, \bibinfo {author} {\bibfnamefont {E.}~\bibnamefont {Tonni}},\ and\
  \bibinfo {author} {\bibfnamefont {I.}~\bibnamefont {Peschel}},\ }\bibfield
  {title} {\bibinfo {title} {Local and non-local properties of the entanglement
  hamiltonian for two disjoint intervals},\ }\href
  {https://doi.org/10.1088/1742-5468/ac8151} {\bibfield  {journal} {\bibinfo
  {journal} {Journal of Statistical Mechanics: Theory and Experiment}\ }\textbf
  {\bibinfo {volume} {2022}},\ \bibinfo {pages} {083101} (\bibinfo {year}
  {2022})}\BibitemShut {NoStop}%
\bibitem [{\citenamefont {Fidkowski}(2010)}]{Fidkowski}%
  \BibitemOpen
  \bibfield  {author} {\bibinfo {author} {\bibfnamefont {L.}~\bibnamefont
  {Fidkowski}},\ }\bibfield  {title} {\bibinfo {title} {Entanglement spectrum
  of topological insulators and superconductors},\ }\href
  {https://doi.org/10.1103/PhysRevLett.104.130502} {\bibfield  {journal}
  {\bibinfo  {journal} {Phys. Rev. Lett.}\ }\textbf {\bibinfo {volume} {104}},\
  \bibinfo {pages} {130502} (\bibinfo {year} {2010})}\BibitemShut {NoStop}%
\bibitem [{\citenamefont {Yates}\ and\ \citenamefont {Mitra}(2017)}]{Aditi}%
  \BibitemOpen
  \bibfield  {author} {\bibinfo {author} {\bibfnamefont {D.~J.}\ \bibnamefont
  {Yates}}\ and\ \bibinfo {author} {\bibfnamefont {A.}~\bibnamefont {Mitra}},\
  }\bibfield  {title} {\bibinfo {title} {Entanglement properties of the
  time-periodic {K}itaev chain},\ }\href
  {https://doi.org/10.1103/PhysRevB.96.115108} {\bibfield  {journal} {\bibinfo
  {journal} {Phys. Rev. B}\ }\textbf {\bibinfo {volume} {96}},\ \bibinfo
  {pages} {115108} (\bibinfo {year} {2017})}\BibitemShut {NoStop}%
\bibitem [{\citenamefont {Arildsen}\ \emph {et~al.}(2024)\citenamefont
  {Arildsen}, \citenamefont {Chen}, \citenamefont {Schuch},\ and\ \citenamefont
  {Ludwig}}]{QSL_chiral}%
  \BibitemOpen
  \bibfield  {author} {\bibinfo {author} {\bibfnamefont {M.~J.}\ \bibnamefont
  {Arildsen}}, \bibinfo {author} {\bibfnamefont {J.-Y.}\ \bibnamefont {Chen}},
  \bibinfo {author} {\bibfnamefont {N.}~\bibnamefont {Schuch}},\ and\ \bibinfo
  {author} {\bibfnamefont {A.~W.~W.}\ \bibnamefont {Ludwig}},\ }\bibfield
  {title} {\bibinfo {title} {Entanglement spectrum as a diagnostic of chirality
  of topological spin liquids: Analysis of {SU(3)} projected entangled pair
  states},\ }\href {https://doi.org/10.1103/PhysRevB.110.235147} {\bibfield
  {journal} {\bibinfo  {journal} {Phys. Rev. B}\ }\textbf {\bibinfo {volume}
  {110}},\ \bibinfo {pages} {235147} (\bibinfo {year} {2024})}\BibitemShut
  {NoStop}%
\bibitem [{\citenamefont {Hasebe}\ and\ \citenamefont
  {Totsuka}(2013)}]{PhysRevB.87.045115}%
  \BibitemOpen
  \bibfield  {author} {\bibinfo {author} {\bibfnamefont {K.}~\bibnamefont
  {Hasebe}}\ and\ \bibinfo {author} {\bibfnamefont {K.}~\bibnamefont
  {Totsuka}},\ }\bibfield  {title} {\bibinfo {title} {Quantum entanglement and
  topological order in hole-doped valence-bond solid states},\ }\href
  {https://doi.org/10.1103/PhysRevB.87.045115} {\bibfield  {journal} {\bibinfo
  {journal} {Phys. Rev. B}\ }\textbf {\bibinfo {volume} {87}},\ \bibinfo
  {pages} {045115} (\bibinfo {year} {2013})}\BibitemShut {NoStop}%
\bibitem [{\citenamefont {Eisler}\ \emph {et~al.}(2009)\citenamefont {Eisler},
  \citenamefont {Iglói},\ and\ \citenamefont {Peschel}}]{Eisler_2009}%
  \BibitemOpen
  \bibfield  {author} {\bibinfo {author} {\bibfnamefont {V.}~\bibnamefont
  {Eisler}}, \bibinfo {author} {\bibfnamefont {F.}~\bibnamefont {Iglói}},\
  and\ \bibinfo {author} {\bibfnamefont {I.}~\bibnamefont {Peschel}},\
  }\bibfield  {title} {\bibinfo {title} {Entanglement in spin chains with
  gradients},\ }\href {https://doi.org/10.1088/1742-5468/2009/02/P02011}
  {\bibfield  {journal} {\bibinfo  {journal} {Journal of Statistical Mechanics:
  Theory and Experiment}\ }\textbf {\bibinfo {volume} {2009}},\ \bibinfo
  {pages} {P02011} (\bibinfo {year} {2009})}\BibitemShut {NoStop}%
\bibitem [{\citenamefont {Zhao}\ \emph {et~al.}(2006)\citenamefont {Zhao},
  \citenamefont {Peschel},\ and\ \citenamefont {Wang}}]{XXZ_EE}%
  \BibitemOpen
  \bibfield  {author} {\bibinfo {author} {\bibfnamefont {J.}~\bibnamefont
  {Zhao}}, \bibinfo {author} {\bibfnamefont {I.}~\bibnamefont {Peschel}},\ and\
  \bibinfo {author} {\bibfnamefont {X.}~\bibnamefont {Wang}},\ }\bibfield
  {title} {\bibinfo {title} {Critical entanglement of {XXZ} heisenberg chains
  with defects},\ }\href {https://doi.org/10.1103/PhysRevB.73.024417}
  {\bibfield  {journal} {\bibinfo  {journal} {Phys. Rev. B}\ }\textbf {\bibinfo
  {volume} {73}},\ \bibinfo {pages} {024417} (\bibinfo {year}
  {2006})}\BibitemShut {NoStop}%
\bibitem [{\citenamefont {Thomale}\ \emph
  {et~al.}(2010{\natexlab{a}})\citenamefont {Thomale}, \citenamefont {Arovas},\
  and\ \citenamefont {Bernevig}}]{Gapless_spin_chain}%
  \BibitemOpen
  \bibfield  {author} {\bibinfo {author} {\bibfnamefont {R.}~\bibnamefont
  {Thomale}}, \bibinfo {author} {\bibfnamefont {D.~P.}\ \bibnamefont
  {Arovas}},\ and\ \bibinfo {author} {\bibfnamefont {B.~A.}\ \bibnamefont
  {Bernevig}},\ }\bibfield  {title} {\bibinfo {title} {Nonlocal order in
  gapless systems: Entanglement spectrum in spin chains},\ }\href
  {https://doi.org/10.1103/PhysRevLett.105.116805} {\bibfield  {journal}
  {\bibinfo  {journal} {Phys. Rev. Lett.}\ }\textbf {\bibinfo {volume} {105}},\
  \bibinfo {pages} {116805} (\bibinfo {year} {2010}{\natexlab{a}})}\BibitemShut
  {NoStop}%
\bibitem [{\citenamefont {Ju}\ \emph {et~al.}(2012)\citenamefont {Ju},
  \citenamefont {Kallin}, \citenamefont {Fendley}, \citenamefont {Hastings},\
  and\ \citenamefont {Melko}}]{2D_gapless}%
  \BibitemOpen
  \bibfield  {author} {\bibinfo {author} {\bibfnamefont {H.}~\bibnamefont
  {Ju}}, \bibinfo {author} {\bibfnamefont {A.~B.}\ \bibnamefont {Kallin}},
  \bibinfo {author} {\bibfnamefont {P.}~\bibnamefont {Fendley}}, \bibinfo
  {author} {\bibfnamefont {M.~B.}\ \bibnamefont {Hastings}},\ and\ \bibinfo
  {author} {\bibfnamefont {R.~G.}\ \bibnamefont {Melko}},\ }\bibfield  {title}
  {\bibinfo {title} {Entanglement scaling in two-dimensional gapless systems},\
  }\href {https://doi.org/10.1103/PhysRevB.85.165121} {\bibfield  {journal}
  {\bibinfo  {journal} {Phys. Rev. B}\ }\textbf {\bibinfo {volume} {85}},\
  \bibinfo {pages} {165121} (\bibinfo {year} {2012})}\BibitemShut {NoStop}%
\bibitem [{\citenamefont {Nakagawa}\ and\ \citenamefont
  {Furukawa}(2017)}]{COE_Shunshuke}%
  \BibitemOpen
  \bibfield  {author} {\bibinfo {author} {\bibfnamefont {Y.~O.}\ \bibnamefont
  {Nakagawa}}\ and\ \bibinfo {author} {\bibfnamefont {S.}~\bibnamefont
  {Furukawa}},\ }\bibfield  {title} {\bibinfo {title} {Capacity of entanglement
  and the distribution of density matrix eigenvalues in gapless systems},\
  }\href {https://doi.org/10.1103/PhysRevB.96.205108} {\bibfield  {journal}
  {\bibinfo  {journal} {Phys. Rev. B}\ }\textbf {\bibinfo {volume} {96}},\
  \bibinfo {pages} {205108} (\bibinfo {year} {2017})}\BibitemShut {NoStop}%
\bibitem [{\citenamefont {Chen}\ \emph {et~al.}(2018)\citenamefont {Chen},
  \citenamefont {Hung}, \citenamefont {Li},\ and\ \citenamefont
  {Wan}}]{JHEP_2018}%
  \BibitemOpen
  \bibfield  {author} {\bibinfo {author} {\bibfnamefont {C.}~\bibnamefont
  {Chen}}, \bibinfo {author} {\bibfnamefont {L.-Y.}\ \bibnamefont {Hung}},
  \bibinfo {author} {\bibfnamefont {Y.}~\bibnamefont {Li}},\ and\ \bibinfo
  {author} {\bibfnamefont {Y.}~\bibnamefont {Wan}},\ }\bibfield  {title}
  {\bibinfo {title} {Entanglement entropy of topological orders with
  boundaries},\ }\href {https://doi.org/10.1007/JHEP06(2018)113} {\bibfield
  {journal} {\bibinfo  {journal} {Journal of High Energy Physics}\ }\textbf
  {\bibinfo {volume} {2018}},\ \bibinfo {pages} {113} (\bibinfo {year}
  {2018})}\BibitemShut {NoStop}%
\bibitem [{HAM(2005)}]{HAMMA200522}%
  \BibitemOpen
  \bibfield  {title} {\bibinfo {title} {Ground state entanglement and geometric
  entropy in the {K}itaev model},\ }\href
  {https://doi.org/https://doi.org/10.1016/j.physleta.2005.01.060} {\bibfield
  {journal} {\bibinfo  {journal} {Physics Letters A}\ }\textbf {\bibinfo
  {volume} {337}},\ \bibinfo {pages} {22} (\bibinfo {year} {2005})}\BibitemShut
  {NoStop}%
\bibitem [{\citenamefont {Hamma}\ \emph {et~al.}(2005)\citenamefont {Hamma},
  \citenamefont {Ionicioiu},\ and\ \citenamefont
  {Zanardi}}]{Hamma_lattice_spin}%
  \BibitemOpen
  \bibfield  {author} {\bibinfo {author} {\bibfnamefont {A.}~\bibnamefont
  {Hamma}}, \bibinfo {author} {\bibfnamefont {R.}~\bibnamefont {Ionicioiu}},\
  and\ \bibinfo {author} {\bibfnamefont {P.}~\bibnamefont {Zanardi}},\
  }\bibfield  {title} {\bibinfo {title} {Bipartite entanglement and entropic
  boundary law in lattice spin systems},\ }\href
  {https://doi.org/10.1103/PhysRevA.71.022315} {\bibfield  {journal} {\bibinfo
  {journal} {Phys. Rev. A}\ }\textbf {\bibinfo {volume} {71}},\ \bibinfo
  {pages} {022315} (\bibinfo {year} {2005})}\BibitemShut {NoStop}%
\bibitem [{\citenamefont {Grover}\ \emph {et~al.}(2013)\citenamefont {Grover},
  \citenamefont {Zhang},\ and\ \citenamefont {Vishwanath}}]{Grover_2013}%
  \BibitemOpen
  \bibfield  {author} {\bibinfo {author} {\bibfnamefont {T.}~\bibnamefont
  {Grover}}, \bibinfo {author} {\bibfnamefont {Y.}~\bibnamefont {Zhang}},\ and\
  \bibinfo {author} {\bibfnamefont {A.}~\bibnamefont {Vishwanath}},\ }\bibfield
   {title} {\bibinfo {title} {Entanglement entropy as a portal to the physics
  of quantum spin liquids},\ }\href
  {https://doi.org/10.1088/1367-2630/15/2/025002} {\bibfield  {journal}
  {\bibinfo  {journal} {New Journal of Physics}\ }\textbf {\bibinfo {volume}
  {15}},\ \bibinfo {pages} {025002} (\bibinfo {year} {2013})}\BibitemShut
  {NoStop}%
\bibitem [{\citenamefont {Shi}\ \emph {et~al.}(2016)\citenamefont {Shi},
  \citenamefont {Wang}, \citenamefont {Li}, \citenamefont {Cho}, \citenamefont
  {Batchelor},\ and\ \citenamefont {Zhou}}]{2D_geometric_entanglement}%
  \BibitemOpen
  \bibfield  {author} {\bibinfo {author} {\bibfnamefont {Q.-Q.}\ \bibnamefont
  {Shi}}, \bibinfo {author} {\bibfnamefont {H.-L.}\ \bibnamefont {Wang}},
  \bibinfo {author} {\bibfnamefont {S.-H.}\ \bibnamefont {Li}}, \bibinfo
  {author} {\bibfnamefont {S.~Y.}\ \bibnamefont {Cho}}, \bibinfo {author}
  {\bibfnamefont {M.~T.}\ \bibnamefont {Batchelor}},\ and\ \bibinfo {author}
  {\bibfnamefont {H.-Q.}\ \bibnamefont {Zhou}},\ }\bibfield  {title} {\bibinfo
  {title} {Geometric entanglement and quantum phase transitions in
  two-dimensional quantum lattice models},\ }\href
  {https://doi.org/10.1103/PhysRevA.93.062341} {\bibfield  {journal} {\bibinfo
  {journal} {Phys. Rev. A}\ }\textbf {\bibinfo {volume} {93}},\ \bibinfo
  {pages} {062341} (\bibinfo {year} {2016})}\BibitemShut {NoStop}%
\bibitem [{\citenamefont {Ye}\ \emph {et~al.}(2022)\citenamefont {Ye},
  \citenamefont {Yang}, \citenamefont {Li},\ and\ \citenamefont {Hu}}]{fphy}%
  \BibitemOpen
  \bibfield  {author} {\bibinfo {author} {\bibfnamefont {D.}~\bibnamefont
  {Ye}}, \bibinfo {author} {\bibfnamefont {Y.}~\bibnamefont {Yang}}, \bibinfo
  {author} {\bibfnamefont {Q.}~\bibnamefont {Li}},\ and\ \bibinfo {author}
  {\bibfnamefont {Z.-X.}\ \bibnamefont {Hu}},\ }\bibfield  {title} {\bibinfo
  {title} {Entanglement entropy of the quantum hall edge and its geometric
  contribution},\ }\href {https://doi.org/10.3389/fphy.2022.971423} {\bibfield
  {journal} {\bibinfo  {journal} {Frontiers in Physics}\ }\textbf {\bibinfo
  {volume} {10}},\ \bibinfo {pages} {971423} (\bibinfo {year}
  {2022})}\BibitemShut {NoStop}%
\bibitem [{\citenamefont {Kumar}\ and\ \citenamefont
  {Bhatt}(2023)}]{Scaling_EE}%
  \BibitemOpen
  \bibfield  {author} {\bibinfo {author} {\bibfnamefont {P.}~\bibnamefont
  {Kumar}}\ and\ \bibinfo {author} {\bibfnamefont {R.~N.}\ \bibnamefont
  {Bhatt}},\ }\bibfield  {title} {\bibinfo {title} {Scaling of entanglement
  entropy at quantum critical points in random spin chains},\ }\href
  {https://doi.org/10.1103/PhysRevB.108.L241113} {\bibfield  {journal}
  {\bibinfo  {journal} {Phys. Rev. B}\ }\textbf {\bibinfo {volume} {108}},\
  \bibinfo {pages} {L241113} (\bibinfo {year} {2023})}\BibitemShut {NoStop}%
\bibitem [{\citenamefont {Kumar}\ \emph {et~al.}(2025)\citenamefont {Kumar},
  \citenamefont {Vedula}, \citenamefont {Gangadharaiah},\ and\ \citenamefont
  {Sharma}}]{EE_probe_AAH}%
  \BibitemOpen
  \bibfield  {author} {\bibinfo {author} {\bibfnamefont {M.}~\bibnamefont
  {Kumar}}, \bibinfo {author} {\bibfnamefont {B.}~\bibnamefont {Vedula}},
  \bibinfo {author} {\bibfnamefont {S.}~\bibnamefont {Gangadharaiah}},\ and\
  \bibinfo {author} {\bibfnamefont {A.}~\bibnamefont {Sharma}},\ }\href
  {https://arxiv.org/abs/2508.15897} {\bibinfo {title} {Entanglement entropy as
  a probe of topological phase transitions}} (\bibinfo {year} {2025}),\ \Eprint
  {https://arxiv.org/abs/2508.15897} {arXiv:2508.15897 [cond-mat.str-el]}
  \BibitemShut {NoStop}%
\bibitem [{\citenamefont {Zhou}\ and\ \citenamefont
  {Ye}(2024)}]{EE_fractalization}%
  \BibitemOpen
  \bibfield  {author} {\bibinfo {author} {\bibfnamefont {Y.}~\bibnamefont
  {Zhou}}\ and\ \bibinfo {author} {\bibfnamefont {P.}~\bibnamefont {Ye}},\
  }\bibfield  {title} {\bibinfo {title} {Entanglement fractalization},\ }\href
  {https://doi.org/10.1103/PhysRevResearch.6.043145} {\bibfield  {journal}
  {\bibinfo  {journal} {Phys. Rev. Res.}\ }\textbf {\bibinfo {volume} {6}},\
  \bibinfo {pages} {043145} (\bibinfo {year} {2024})}\BibitemShut {NoStop}%
\bibitem [{\citenamefont {Pirmoradian}\ and\ \citenamefont
  {Tanhayi}(2024)}]{Pirmoradian2024}%
  \BibitemOpen
  \bibfield  {author} {\bibinfo {author} {\bibfnamefont {R.}~\bibnamefont
  {Pirmoradian}}\ and\ \bibinfo {author} {\bibfnamefont {M.~R.}\ \bibnamefont
  {Tanhayi}},\ }\bibfield  {title} {\bibinfo {title} {Symmetry-resolved
  entanglement entropy for local and non-local qfts},\ }\href
  {https://doi.org/10.1140/epjc/s10052-024-13212-8} {\bibfield  {journal}
  {\bibinfo  {journal} {The European Physical Journal C}\ }\textbf {\bibinfo
  {volume} {84}},\ \bibinfo {pages} {849} (\bibinfo {year} {2024})}\BibitemShut
  {NoStop}%
\bibitem [{\citenamefont {Song}\ \emph {et~al.}(2024)\citenamefont {Song},
  \citenamefont {Zhao}, \citenamefont {Meng}, \citenamefont {Xu},\ and\
  \citenamefont {Cheng}}]{SciPostPhys.17.1.010}%
  \BibitemOpen
  \bibfield  {author} {\bibinfo {author} {\bibfnamefont {M.}~\bibnamefont
  {Song}}, \bibinfo {author} {\bibfnamefont {J.}~\bibnamefont {Zhao}}, \bibinfo
  {author} {\bibfnamefont {Z.~Y.}\ \bibnamefont {Meng}}, \bibinfo {author}
  {\bibfnamefont {C.}~\bibnamefont {Xu}},\ and\ \bibinfo {author}
  {\bibfnamefont {M.}~\bibnamefont {Cheng}},\ }\bibfield  {title} {\bibinfo
  {title} {{Extracting subleading corrections in entanglement entropy at
  quantum phase transitions}},\ }\href
  {https://doi.org/10.21468/SciPostPhys.17.1.010} {\bibfield  {journal}
  {\bibinfo  {journal} {SciPost Phys.}\ }\textbf {\bibinfo {volume} {17}},\
  \bibinfo {pages} {010} (\bibinfo {year} {2024})}\BibitemShut {NoStop}%
\bibitem [{\citenamefont {Nehra}\ \emph {et~al.}(2020)\citenamefont {Nehra},
  \citenamefont {Bhakuni}, \citenamefont {Ramachandran},\ and\ \citenamefont
  {Sharma}}]{Kitaev_Ladder}%
  \BibitemOpen
  \bibfield  {author} {\bibinfo {author} {\bibfnamefont {R.}~\bibnamefont
  {Nehra}}, \bibinfo {author} {\bibfnamefont {D.~S.}\ \bibnamefont {Bhakuni}},
  \bibinfo {author} {\bibfnamefont {A.}~\bibnamefont {Ramachandran}},\ and\
  \bibinfo {author} {\bibfnamefont {A.}~\bibnamefont {Sharma}},\ }\bibfield
  {title} {\bibinfo {title} {Flat bands and entanglement in the {K}itaev
  ladder},\ }\href {https://doi.org/10.1103/PhysRevResearch.2.013175}
  {\bibfield  {journal} {\bibinfo  {journal} {Phys. Rev. Res.}\ }\textbf
  {\bibinfo {volume} {2}},\ \bibinfo {pages} {013175} (\bibinfo {year}
  {2020})}\BibitemShut {NoStop}%
\bibitem [{\citenamefont {Guo}\ \emph {et~al.}(2025)\citenamefont {Guo},
  \citenamefont {Yang},\ and\ \citenamefont {Yu}}]{Critical_Free_fermion}%
  \BibitemOpen
  \bibfield  {author} {\bibinfo {author} {\bibfnamefont {Y.}~\bibnamefont
  {Guo}}, \bibinfo {author} {\bibfnamefont {S.}~\bibnamefont {Yang}},\ and\
  \bibinfo {author} {\bibfnamefont {X.-J.}\ \bibnamefont {Yu}},\ }\href
  {https://arxiv.org/abs/2509.20054} {\bibinfo {title} {Generalized li-haldane
  correspondence in critical free-fermion systems}} (\bibinfo {year} {2025}),\
  \Eprint {https://arxiv.org/abs/2509.20054} {arXiv:2509.20054
  [cond-mat.str-el]} \BibitemShut {NoStop}%
\bibitem [{\citenamefont {Haim}\ and\ \citenamefont {Oreg}(2019)}]{TRI_review}%
  \BibitemOpen
  \bibfield  {author} {\bibinfo {author} {\bibfnamefont {A.}~\bibnamefont
  {Haim}}\ and\ \bibinfo {author} {\bibfnamefont {Y.}~\bibnamefont {Oreg}},\
  }\bibfield  {title} {\bibinfo {title} {Time-reversal-invariant topological
  superconductivity in one and two dimensions},\ }\href
  {https://doi.org/https://doi.org/10.1016/j.physrep.2019.08.002} {\bibfield
  {journal} {\bibinfo  {journal} {Physics Reports}\ }\textbf {\bibinfo {volume}
  {825}},\ \bibinfo {pages} {1} (\bibinfo {year} {2019})}\BibitemShut {NoStop}%
\bibitem [{\citenamefont {Chinellato}\ \emph {et~al.}(2024)\citenamefont
  {Chinellato}, \citenamefont {Gazza}, \citenamefont {Lobos},\ and\
  \citenamefont {Aligia}}]{TRI_1}%
  \BibitemOpen
  \bibfield  {author} {\bibinfo {author} {\bibfnamefont {L.~M.}\ \bibnamefont
  {Chinellato}}, \bibinfo {author} {\bibfnamefont {C.~J.}\ \bibnamefont
  {Gazza}}, \bibinfo {author} {\bibfnamefont {A.~M.}\ \bibnamefont {Lobos}},\
  and\ \bibinfo {author} {\bibfnamefont {A.~A.}\ \bibnamefont {Aligia}},\
  }\bibfield  {title} {\bibinfo {title} {Topological phases of strongly
  interacting time-reversal invariant topological superconducting chains under
  a magnetic field},\ }\href {https://doi.org/10.1103/PhysRevB.109.064503}
  {\bibfield  {journal} {\bibinfo  {journal} {Phys. Rev. B}\ }\textbf {\bibinfo
  {volume} {109}},\ \bibinfo {pages} {064503} (\bibinfo {year}
  {2024})}\BibitemShut {NoStop}%
\bibitem [{\citenamefont {Haim}\ \emph {et~al.}(2014)\citenamefont {Haim},
  \citenamefont {Keselman}, \citenamefont {Berg},\ and\ \citenamefont
  {Oreg}}]{TRI_2}%
  \BibitemOpen
  \bibfield  {author} {\bibinfo {author} {\bibfnamefont {A.}~\bibnamefont
  {Haim}}, \bibinfo {author} {\bibfnamefont {A.}~\bibnamefont {Keselman}},
  \bibinfo {author} {\bibfnamefont {E.}~\bibnamefont {Berg}},\ and\ \bibinfo
  {author} {\bibfnamefont {Y.}~\bibnamefont {Oreg}},\ }\bibfield  {title}
  {\bibinfo {title} {Time-reversal-invariant topological superconductivity
  induced by repulsive interactions in quantum wires},\ }\href
  {https://doi.org/10.1103/PhysRevB.89.220504} {\bibfield  {journal} {\bibinfo
  {journal} {Phys. Rev. B}\ }\textbf {\bibinfo {volume} {89}},\ \bibinfo
  {pages} {220504} (\bibinfo {year} {2014})}\BibitemShut {NoStop}%
\bibitem [{\citenamefont {Haim}\ \emph {et~al.}(2016)\citenamefont {Haim},
  \citenamefont {Berg}, \citenamefont {Flensberg},\ and\ \citenamefont
  {Oreg}}]{TRI_3}%
  \BibitemOpen
  \bibfield  {author} {\bibinfo {author} {\bibfnamefont {A.}~\bibnamefont
  {Haim}}, \bibinfo {author} {\bibfnamefont {E.}~\bibnamefont {Berg}}, \bibinfo
  {author} {\bibfnamefont {K.}~\bibnamefont {Flensberg}},\ and\ \bibinfo
  {author} {\bibfnamefont {Y.}~\bibnamefont {Oreg}},\ }\bibfield  {title}
  {\bibinfo {title} {No-go theorem for a time-reversal invariant topological
  phase in noninteracting systems coupled to conventional superconductors},\
  }\href {https://doi.org/10.1103/PhysRevB.94.161110} {\bibfield  {journal}
  {\bibinfo  {journal} {Phys. Rev. B}\ }\textbf {\bibinfo {volume} {94}},\
  \bibinfo {pages} {161110} (\bibinfo {year} {2016})}\BibitemShut {NoStop}%
\bibitem [{\citenamefont {Schrade}\ and\ \citenamefont
  {Fu}(2022)}]{QC_Majorana_Kramers}%
  \BibitemOpen
  \bibfield  {author} {\bibinfo {author} {\bibfnamefont {C.}~\bibnamefont
  {Schrade}}\ and\ \bibinfo {author} {\bibfnamefont {L.}~\bibnamefont {Fu}},\
  }\bibfield  {title} {\bibinfo {title} {Quantum computing with majorana
  kramers pairs},\ }\href {https://doi.org/10.1103/PhysRevLett.129.227002}
  {\bibfield  {journal} {\bibinfo  {journal} {Phys. Rev. Lett.}\ }\textbf
  {\bibinfo {volume} {129}},\ \bibinfo {pages} {227002} (\bibinfo {year}
  {2022})}\BibitemShut {NoStop}%
\bibitem [{\citenamefont {Thomale}\ \emph
  {et~al.}(2010{\natexlab{b}})\citenamefont {Thomale}, \citenamefont {Arovas},\
  and\ \citenamefont {Bernevig}}]{GSPT_Bernevig}%
  \BibitemOpen
  \bibfield  {author} {\bibinfo {author} {\bibfnamefont {R.}~\bibnamefont
  {Thomale}}, \bibinfo {author} {\bibfnamefont {D.~P.}\ \bibnamefont
  {Arovas}},\ and\ \bibinfo {author} {\bibfnamefont {B.~A.}\ \bibnamefont
  {Bernevig}},\ }\bibfield  {title} {\bibinfo {title} {Nonlocal order in
  gapless systems: Entanglement spectrum in spin chains},\ }\href
  {https://doi.org/10.1103/PhysRevLett.105.116805} {\bibfield  {journal}
  {\bibinfo  {journal} {Phys. Rev. Lett.}\ }\textbf {\bibinfo {volume} {105}},\
  \bibinfo {pages} {116805} (\bibinfo {year} {2010}{\natexlab{b}})}\BibitemShut
  {NoStop}%
\bibitem [{\citenamefont {Komijani}(2020)}]{Kondo_anyons_TQC}%
  \BibitemOpen
  \bibfield  {author} {\bibinfo {author} {\bibfnamefont {Y.}~\bibnamefont
  {Komijani}},\ }\bibfield  {title} {\bibinfo {title} {Isolating kondo anyons
  for topological quantum computation},\ }\href
  {https://doi.org/10.1103/PhysRevB.101.235131} {\bibfield  {journal} {\bibinfo
   {journal} {Phys. Rev. B}\ }\textbf {\bibinfo {volume} {101}},\ \bibinfo
  {pages} {235131} (\bibinfo {year} {2020})}\BibitemShut {NoStop}%
\bibitem [{\citenamefont {Chatterjee}\ \emph {et~al.}(2024)\citenamefont
  {Chatterjee}, \citenamefont {Banik}, \citenamefont {Bera}, \citenamefont
  {Ghosh}, \citenamefont {Pradhan}, \citenamefont {Saha},\ and\ \citenamefont
  {Nandy}}]{exp_realisation_Arijit}%
  \BibitemOpen
  \bibfield  {author} {\bibinfo {author} {\bibfnamefont {P.}~\bibnamefont
  {Chatterjee}}, \bibinfo {author} {\bibfnamefont {S.}~\bibnamefont {Banik}},
  \bibinfo {author} {\bibfnamefont {S.}~\bibnamefont {Bera}}, \bibinfo {author}
  {\bibfnamefont {A.~K.}\ \bibnamefont {Ghosh}}, \bibinfo {author}
  {\bibfnamefont {S.}~\bibnamefont {Pradhan}}, \bibinfo {author} {\bibfnamefont
  {A.}~\bibnamefont {Saha}},\ and\ \bibinfo {author} {\bibfnamefont {A.~K.}\
  \bibnamefont {Nandy}},\ }\bibfield  {title} {\bibinfo {title} {Topological
  superconductivity by engineering noncollinear magnetism in
  magnet/superconductor heterostructures: A realistic prescription for the
  two-dimensional {K}itaev model},\ }\href
  {https://doi.org/10.1103/PhysRevB.109.L121301} {\bibfield  {journal}
  {\bibinfo  {journal} {Phys. Rev. B}\ }\textbf {\bibinfo {volume} {109}},\
  \bibinfo {pages} {L121301} (\bibinfo {year} {2024})}\BibitemShut {NoStop}%
\bibitem [{\citenamefont {{K}itaev}(2001)}]{Kitaev}%
  \BibitemOpen
  \bibfield  {author} {\bibinfo {author} {\bibfnamefont {A.~Y.}\ \bibnamefont
  {{K}itaev}},\ }\bibfield  {title} {\bibinfo {title} {Unpaired {M}ajorana
  fermions in quantum wires},\ }\href
  {https://doi.org/10.1070/1063-7869/44/10S/S29} {\bibfield  {journal}
  {\bibinfo  {journal} {Physics-Uspekhi}\ }\textbf {\bibinfo {volume} {44}},\
  \bibinfo {pages} {131} (\bibinfo {year} {2001})}\BibitemShut {NoStop}%
\bibitem [{\citenamefont {Lieb}\ \emph {et~al.}(1961)\citenamefont {Lieb},
  \citenamefont {Schultz},\ and\ \citenamefont {Mattis}}]{LSM}%
  \BibitemOpen
  \bibfield  {author} {\bibinfo {author} {\bibfnamefont {E.}~\bibnamefont
  {Lieb}}, \bibinfo {author} {\bibfnamefont {T.}~\bibnamefont {Schultz}},\ and\
  \bibinfo {author} {\bibfnamefont {D.}~\bibnamefont {Mattis}},\ }\bibfield
  {title} {\bibinfo {title} {Two soluble models of an antiferromagnetic
  chain},\ }\href
  {https://doi.org/https://doi.org/10.1016/0003-4916(61)90115-4} {\bibfield
  {journal} {\bibinfo  {journal} {Annals of Physics}\ }\textbf {\bibinfo
  {volume} {16}},\ \bibinfo {pages} {407} (\bibinfo {year} {1961})}\BibitemShut
  {NoStop}%
\bibitem [{\citenamefont {Mahyaeh}\ and\ \citenamefont {Ardonne}(2018)}]{NJP}%
  \BibitemOpen
  \bibfield  {author} {\bibinfo {author} {\bibfnamefont {I.}~\bibnamefont
  {Mahyaeh}}\ and\ \bibinfo {author} {\bibfnamefont {E.}~\bibnamefont
  {Ardonne}},\ }\bibfield  {title} {\bibinfo {title} {Zero modes of the
  {K}itaev chain with phase-gradients and longer range couplings},\ }\href
  {https://doi.org/10.1088/2399-6528/aab7e5} {\bibfield  {journal} {\bibinfo
  {journal} {Journal of Physics Communications}\ }\textbf {\bibinfo {volume}
  {2}},\ \bibinfo {pages} {045010} (\bibinfo {year} {2018})}\BibitemShut
  {NoStop}%
\bibitem [{\citenamefont {Li}\ and\ \citenamefont
  {Haldane}(2008)}]{Li_Haldane_2008}%
  \BibitemOpen
  \bibfield  {author} {\bibinfo {author} {\bibfnamefont {H.}~\bibnamefont
  {Li}}\ and\ \bibinfo {author} {\bibfnamefont {F.~D.~M.}\ \bibnamefont
  {Haldane}},\ }\bibfield  {title} {\bibinfo {title} {Entanglement spectrum as
  a generalization of entanglement entropy: Identification of topological order
  in non-abelian fractional quantum hall effect states},\ }\href
  {https://doi.org/10.1103/PhysRevLett.101.010504} {\bibfield  {journal}
  {\bibinfo  {journal} {Phys. Rev. Lett.}\ }\textbf {\bibinfo {volume} {101}},\
  \bibinfo {pages} {010504} (\bibinfo {year} {2008})}\BibitemShut {NoStop}%
\bibitem [{\citenamefont {Iizuka}\ \emph {et~al.}(2023)\citenamefont {Iizuka},
  \citenamefont {Yuan}, \citenamefont {Mita}, \citenamefont {Higo},
  \citenamefont {Yasunaga},\ and\ \citenamefont {Ezawa}}]{top_circuits}%
  \BibitemOpen
  \bibfield  {author} {\bibinfo {author} {\bibfnamefont {T.}~\bibnamefont
  {Iizuka}}, \bibinfo {author} {\bibfnamefont {H.}~\bibnamefont {Yuan}},
  \bibinfo {author} {\bibfnamefont {Y.}~\bibnamefont {Mita}}, \bibinfo {author}
  {\bibfnamefont {A.}~\bibnamefont {Higo}}, \bibinfo {author} {\bibfnamefont
  {S.}~\bibnamefont {Yasunaga}},\ and\ \bibinfo {author} {\bibfnamefont
  {M.}~\bibnamefont {Ezawa}},\ }\bibfield  {title} {\bibinfo {title}
  {Experimental demonstration of position-controllable topological interface
  states in high-frequency {K}itaev topological integrated circuits},\ }\href
  {https://doi.org/10.1038/s42005-023-01404-9} {\bibfield  {journal} {\bibinfo
  {journal} {Communications Physics}\ }\textbf {\bibinfo {volume} {6}},\
  \bibinfo {pages} {279} (\bibinfo {year} {2023})}\BibitemShut {NoStop}%
\bibitem [{\citenamefont {Zhang}\ \emph {et~al.}(2023)\citenamefont {Zhang},
  \citenamefont {He}, \citenamefont {Sun}, \citenamefont {Zheng}, \citenamefont
  {Liu}, \citenamefont {Luo}, \citenamefont {Wang}, \citenamefont {Zhu},
  \citenamefont {Qiu}, \citenamefont {Shen}, \citenamefont {Wang},
  \citenamefont {Lin}, \citenamefont {Yu}, \citenamefont {Li}, \citenamefont
  {Xiao}, \citenamefont {Li}, \citenamefont {Yang}, \citenamefont {Jiang},
  \citenamefont {Dai}, \citenamefont {Zhou}, \citenamefont {Ma}, \citenamefont
  {Yuan},\ and\ \citenamefont {Pan}}]{EE_optical_lattice}%
  \BibitemOpen
  \bibfield  {author} {\bibinfo {author} {\bibfnamefont {W.-Y.}\ \bibnamefont
  {Zhang}}, \bibinfo {author} {\bibfnamefont {M.-G.}\ \bibnamefont {He}},
  \bibinfo {author} {\bibfnamefont {H.}~\bibnamefont {Sun}}, \bibinfo {author}
  {\bibfnamefont {Y.-G.}\ \bibnamefont {Zheng}}, \bibinfo {author}
  {\bibfnamefont {Y.}~\bibnamefont {Liu}}, \bibinfo {author} {\bibfnamefont
  {A.}~\bibnamefont {Luo}}, \bibinfo {author} {\bibfnamefont {H.-Y.}\
  \bibnamefont {Wang}}, \bibinfo {author} {\bibfnamefont {Z.-H.}\ \bibnamefont
  {Zhu}}, \bibinfo {author} {\bibfnamefont {P.-Y.}\ \bibnamefont {Qiu}},
  \bibinfo {author} {\bibfnamefont {Y.-C.}\ \bibnamefont {Shen}}, \bibinfo
  {author} {\bibfnamefont {X.-K.}\ \bibnamefont {Wang}}, \bibinfo {author}
  {\bibfnamefont {W.}~\bibnamefont {Lin}}, \bibinfo {author} {\bibfnamefont
  {S.-T.}\ \bibnamefont {Yu}}, \bibinfo {author} {\bibfnamefont {B.-C.}\
  \bibnamefont {Li}}, \bibinfo {author} {\bibfnamefont {B.}~\bibnamefont
  {Xiao}}, \bibinfo {author} {\bibfnamefont {M.-D.}\ \bibnamefont {Li}},
  \bibinfo {author} {\bibfnamefont {Y.-M.}\ \bibnamefont {Yang}}, \bibinfo
  {author} {\bibfnamefont {X.}~\bibnamefont {Jiang}}, \bibinfo {author}
  {\bibfnamefont {H.-N.}\ \bibnamefont {Dai}}, \bibinfo {author} {\bibfnamefont
  {Y.}~\bibnamefont {Zhou}}, \bibinfo {author} {\bibfnamefont {X.}~\bibnamefont
  {Ma}}, \bibinfo {author} {\bibfnamefont {Z.-S.}\ \bibnamefont {Yuan}},\ and\
  \bibinfo {author} {\bibfnamefont {J.-W.}\ \bibnamefont {Pan}},\ }\bibfield
  {title} {\bibinfo {title} {Scalable multipartite entanglement created by spin
  exchange in an optical lattice},\ }\href
  {https://doi.org/10.1103/PhysRevLett.131.073401} {\bibfield  {journal}
  {\bibinfo  {journal} {Phys. Rev. Lett.}\ }\textbf {\bibinfo {volume} {131}},\
  \bibinfo {pages} {073401} (\bibinfo {year} {2023})}\BibitemShut {NoStop}%
\bibitem [{\citenamefont {Lin}\ \emph {et~al.}(2024)\citenamefont {Lin},
  \citenamefont {Zhou}, \citenamefont {Jiang}, \citenamefont {Wu},
  \citenamefont {Chen}, \citenamefont {Liu}, \citenamefont {Wang},
  \citenamefont {Ye},\ and\ \citenamefont {Jiang}}]{EE_phononic_system}%
  \BibitemOpen
  \bibfield  {author} {\bibinfo {author} {\bibfnamefont {Z.-K.}\ \bibnamefont
  {Lin}}, \bibinfo {author} {\bibfnamefont {Y.}~\bibnamefont {Zhou}}, \bibinfo
  {author} {\bibfnamefont {B.}~\bibnamefont {Jiang}}, \bibinfo {author}
  {\bibfnamefont {B.-Q.}\ \bibnamefont {Wu}}, \bibinfo {author} {\bibfnamefont
  {L.-M.}\ \bibnamefont {Chen}}, \bibinfo {author} {\bibfnamefont {X.-Y.}\
  \bibnamefont {Liu}}, \bibinfo {author} {\bibfnamefont {L.-W.}\ \bibnamefont
  {Wang}}, \bibinfo {author} {\bibfnamefont {P.}~\bibnamefont {Ye}},\ and\
  \bibinfo {author} {\bibfnamefont {J.-H.}\ \bibnamefont {Jiang}},\ }\bibfield
  {title} {\bibinfo {title} {Measuring entanglement entropy and its topological
  signature for phononic systems},\ }\href
  {https://doi.org/10.1038/s41467-024-45887-8} {\bibfield  {journal} {\bibinfo
  {journal} {Nature Communications}\ }\textbf {\bibinfo {volume} {15}},\
  \bibinfo {pages} {1601} (\bibinfo {year} {2024})}\BibitemShut {NoStop}%
\bibitem [{\citenamefont {Chen}(2016)}]{Chen_2016}%
  \BibitemOpen
  \bibfield  {author} {\bibinfo {author} {\bibfnamefont {W.}~\bibnamefont
  {Chen}},\ }\bibfield  {title} {\bibinfo {title} {Scaling theory of
  topological phase transitions},\ }\href
  {https://doi.org/10.1088/0953-8984/28/5/055601} {\bibfield  {journal}
  {\bibinfo  {journal} {Journal of Physics: Condensed Matter}\ }\textbf
  {\bibinfo {volume} {28}},\ \bibinfo {pages} {055601} (\bibinfo {year}
  {2016})}\BibitemShut {NoStop}%
\bibitem [{\citenamefont {Chen}\ \emph {et~al.}(2016)\citenamefont {Chen},
  \citenamefont {Sigrist},\ and\ \citenamefont {Schnyder}}]{WChen_2016}%
  \BibitemOpen
  \bibfield  {author} {\bibinfo {author} {\bibfnamefont {W.}~\bibnamefont
  {Chen}}, \bibinfo {author} {\bibfnamefont {M.}~\bibnamefont {Sigrist}},\ and\
  \bibinfo {author} {\bibfnamefont {A.~P.}\ \bibnamefont {Schnyder}},\
  }\bibfield  {title} {\bibinfo {title} {Scaling theory of z2 topological
  invariants},\ }\href {https://doi.org/10.1088/0953-8984/28/36/365501}
  {\bibfield  {journal} {\bibinfo  {journal} {Journal of Physics: Condensed
  Matter}\ }\textbf {\bibinfo {volume} {28}},\ \bibinfo {pages} {365501}
  (\bibinfo {year} {2016})}\BibitemShut {NoStop}%
\bibitem [{\citenamefont {Hasan}\ \emph {et~al.}(2021)\citenamefont {Hasan},
  \citenamefont {Chang}, \citenamefont {Belopolski}, \citenamefont {Bian},
  \citenamefont {Xu},\ and\ \citenamefont {Yin}}]{Hasan2021-qz}%
  \BibitemOpen
  \bibfield  {author} {\bibinfo {author} {\bibfnamefont {M.~Z.}\ \bibnamefont
  {Hasan}}, \bibinfo {author} {\bibfnamefont {G.}~\bibnamefont {Chang}},
  \bibinfo {author} {\bibfnamefont {I.}~\bibnamefont {Belopolski}}, \bibinfo
  {author} {\bibfnamefont {G.}~\bibnamefont {Bian}}, \bibinfo {author}
  {\bibfnamefont {S.-Y.}\ \bibnamefont {Xu}},\ and\ \bibinfo {author}
  {\bibfnamefont {J.-X.}\ \bibnamefont {Yin}},\ }\bibfield  {title} {\bibinfo
  {title} {Weyl, {D}irac and high-fold chiral fermions in topological quantum
  matter},\ }\href@noop {} {\bibfield  {journal} {\bibinfo  {journal} {Nat.
  Rev. Mater.}\ }\textbf {\bibinfo {volume} {6}},\ \bibinfo {pages} {784}
  (\bibinfo {year} {2021})}\BibitemShut {NoStop}%
\bibitem [{\citenamefont {Jia}\ \emph {et~al.}(2016)\citenamefont {Jia},
  \citenamefont {Xu},\ and\ \citenamefont {Hasan}}]{Jia2016-qo}%
  \BibitemOpen
  \bibfield  {author} {\bibinfo {author} {\bibfnamefont {S.}~\bibnamefont
  {Jia}}, \bibinfo {author} {\bibfnamefont {S.-Y.}\ \bibnamefont {Xu}},\ and\
  \bibinfo {author} {\bibfnamefont {M.~Z.}\ \bibnamefont {Hasan}},\ }\bibfield
  {title} {\bibinfo {title} {Weyl semimetals, fermi arcs and chiral
  anomalies},\ }\href@noop {} {\bibfield  {journal} {\bibinfo  {journal} {Nat.
  Mater.}\ }\textbf {\bibinfo {volume} {15}},\ \bibinfo {pages} {1140}
  (\bibinfo {year} {2016})}\BibitemShut {NoStop}%
\bibitem [{\citenamefont {F\"unfhaus}\ \emph {et~al.}(2022)\citenamefont
  {F\"unfhaus}, \citenamefont {Kopp},\ and\ \citenamefont
  {Lettl}}]{Funfhaus_2022}%
  \BibitemOpen
  \bibfield  {author} {\bibinfo {author} {\bibfnamefont {A.}~\bibnamefont
  {F\"unfhaus}}, \bibinfo {author} {\bibfnamefont {T.}~\bibnamefont {Kopp}},\
  and\ \bibinfo {author} {\bibfnamefont {E.}~\bibnamefont {Lettl}},\ }\bibfield
   {title} {\bibinfo {title} {Winding vectors of topological defects: multiband
  chern numbers},\ }\href {https://doi.org/10.1088/1751-8121/ac8ef7} {\bibfield
   {journal} {\bibinfo  {journal} {Journal of Physics A: Mathematical and
  Theoretical}\ }\textbf {\bibinfo {volume} {55}},\ \bibinfo {pages} {405202}
  (\bibinfo {year} {2022})}\BibitemShut {NoStop}%
\bibitem [{\citenamefont {Altland}\ and\ \citenamefont
  {Zirnbauer}(1997)}]{symmetry_class1}%
  \BibitemOpen
  \bibfield  {author} {\bibinfo {author} {\bibfnamefont {A.}~\bibnamefont
  {Altland}}\ and\ \bibinfo {author} {\bibfnamefont {M.~R.}\ \bibnamefont
  {Zirnbauer}},\ }\bibfield  {title} {\bibinfo {title} {Nonstandard symmetry
  classes in mesoscopic normal-superconducting hybrid structures},\ }\href
  {https://doi.org/10.1103/PhysRevB.55.1142} {\bibfield  {journal} {\bibinfo
  {journal} {Phys. Rev. B}\ }\textbf {\bibinfo {volume} {55}},\ \bibinfo
  {pages} {1142} (\bibinfo {year} {1997})}\BibitemShut {NoStop}%
\bibitem [{\citenamefont {Schnyder}\ \emph {et~al.}(2008)\citenamefont
  {Schnyder}, \citenamefont {Ryu}, \citenamefont {Furusaki},\ and\
  \citenamefont {Ludwig}}]{symmetry_class2}%
  \BibitemOpen
  \bibfield  {author} {\bibinfo {author} {\bibfnamefont {A.~P.}\ \bibnamefont
  {Schnyder}}, \bibinfo {author} {\bibfnamefont {S.}~\bibnamefont {Ryu}},
  \bibinfo {author} {\bibfnamefont {A.}~\bibnamefont {Furusaki}},\ and\
  \bibinfo {author} {\bibfnamefont {A.~W.~W.}\ \bibnamefont {Ludwig}},\
  }\bibfield  {title} {\bibinfo {title} {Classification of topological
  insulators and superconductors in three spatial dimensions},\ }\href
  {https://doi.org/10.1103/PhysRevB.78.195125} {\bibfield  {journal} {\bibinfo
  {journal} {Phys. Rev. B}\ }\textbf {\bibinfo {volume} {78}},\ \bibinfo
  {pages} {195125} (\bibinfo {year} {2008})}\BibitemShut {NoStop}%
\bibitem [{\citenamefont {Ryu}\ \emph {et~al.}(2010)\citenamefont {Ryu},
  \citenamefont {Schnyder}, \citenamefont {Furusaki},\ and\ \citenamefont
  {Ludwig}}]{symmetry_class3}%
  \BibitemOpen
  \bibfield  {author} {\bibinfo {author} {\bibfnamefont {S.}~\bibnamefont
  {Ryu}}, \bibinfo {author} {\bibfnamefont {A.~P.}\ \bibnamefont {Schnyder}},
  \bibinfo {author} {\bibfnamefont {A.}~\bibnamefont {Furusaki}},\ and\
  \bibinfo {author} {\bibfnamefont {A.~W.~W.}\ \bibnamefont {Ludwig}},\
  }\bibfield  {title} {\bibinfo {title} {Topological insulators and
  superconductors: tenfold way and dimensional hierarchy},\ }\href
  {https://doi.org/10.1088/1367-2630/12/6/065010} {\bibfield  {journal}
  {\bibinfo  {journal} {New Journal of Physics}\ }\textbf {\bibinfo {volume}
  {12}},\ \bibinfo {pages} {065010} (\bibinfo {year} {2010})}\BibitemShut
  {NoStop}%
\end{thebibliography}%

\end{document}